\documentclass[
]{ceurart}

\usepackage{listings}
\usepackage{times}
\usepackage{hyperref}
\usepackage{enumitem}
\usepackage{graphicx}
\usepackage{amsmath}
\usepackage{amssymb}
\usepackage{listings}
\usepackage{color}
\usepackage{url}
\usepackage{geometry}
\usepackage{subcaption} 
\usepackage{tikz}
\usetikzlibrary{positioning, arrows.meta}

\usepackage{booktabs}
\usepackage{tabularx}
\usepackage{longtable}
\usepackage{caption}
\usepackage{array}

\begin{document}

\copyrightyear{2022}
\copyrightclause{Copyright for this paper by its authors.
  Use permitted under Creative Commons License Attribution 4.0
  International (CC BY 4.0).}

\conference{Overlay2025: 7th International Workshop on
Artificial Intelligence and
fOrmal VERification, Logic,
Automata, and sYnthesis}

\title{Leveraging LLMs for Formal Software Requirements: Challenges and Prospects}


\author[1]{Arshad Beg}[%
orcid=0009-0004-6939-0411,
email=arshad.beg@mu.ie,
]
\cormark[1]
\fnmark[1]

\author[1]{Diarmuid O'Donoghue}[%
orcid=0000-0002-3680-4217,
email=diarmuid.odonoghue@mu.ie,
]
\fnmark[1]

\author[1]{Rosemary Monahan}[%
orcid=0000-0003-3886-4675,
email=rosemary.monahan@mu.ie,
]
\fnmark[1]

\address[1]{Maynooth University, Maynooth, Ireland.}

\cortext[1]{Corresponding author.}
\fntext[1]{These authors contributed equally.}


\begin{abstract}
Software correctness is ensured mathematically through formal verification, which requires the generation of a formal requirement specification and an implementation that must be verified. Tools such as model-checkers and theorem provers ensure software correctness by verifying the implementation against the specification. Formal methods deployment is regularly enforced in the development of safety-critical systems e.g. aerospace, medical devices and autonomous systems. Generating these specifications from informal and ambiguous natural language requirements remains the key challenge. Our project, VERIFYAI \footnote{\url{https://github.com/arshadbeg/OVERLAY2025_SupportingDocs.git}}, aims to investigate automated and semi-automated approaches to bridge this gap, using techniques from Natural Language Processing (NLP), ontology-based domain modelling, artefact reuse, and large language models (LLMs). This position paper presents a preliminary synthesis of relevant literature to identify recurring challenges and prospective research directions in the generation of verifiable specifications from informal requirements. 
\end{abstract}
\maketitle

\section{Introduction}

As software systems grow in complexity and criticality, so does the need for scalable verification methods that ensure correctness and reliability. Formal verification is especially important in safety-critical systems where minor software errors can lead to serious consequences, including loss of life, environmental damage, or large-scale system failures. Under all circumstances, the software used in sectors like aviation, healthcare, automotive, and nuclear control must behave exactly as intended. While testing can only check specific scenarios, formal verification uses mathematical techniques to prove that a system meets its specifications under every possible condition.   
This level of assurance is crucial when human safety depends on software behaving reliably. The purpose of sound software engineering principles is to catch flaws early in the design phase, ensuring consistency between requirements and implementation. The mathematical techniques used in formal methods improve trust, and support compliance with industry standards and regulations. However, their adoption in industry is consistently hindered by the challenges of writing and maintaining formal specifications, which demand rigorous developer training and significantly increase the software development cycle time by up to 30\% \cite{Huisman2024}. This motivates research into automated and semi-automated approaches that can make formal verification more accessible to a wider audience of software engineers. 

The VERIFYAI project \cite{Beg2025} aims to address a central challenge of formal software engineering: translating informal, natural-language requirements into formal, verifiable specifications. This paper outlines the challenges and prospects in leveraging Large Language Models (LLMs) for formalising software requirements. The project aims to integrate techniques from Natural Language Processing (NLP), ontology-driven domain modelling, artefact reuse, and large language models (LLMs) to support the automated generation and traceability of formal specifications. Like all other fields, the development of large language models (LLMs) has opened a world of opportunities for the challenge of formalisation of software requirements. We consolidate initial findings to highlight research gaps and recurring difficulties in LLM-assisted formal specification generation. Noting firstly that the main body of the paper is supported by several detailed Appendices available at \url{https://github.com/arshadbeg/OVERLAY2025_SupportingDocs.git}.

\noindent{The key contributions of this paper are as follows:}

\noindent\textbf{Identification of Core Challenges in Formalisation:} We present a structured analysis of the barriers to translating informal, natural language software requirements into formal specifications, such as ambiguity, lack of domain models, and LLM instability.

\noindent\textbf{VERIFYAI Research Framework:} We propose our research framework, which integrates LLMs with NLP, ontology-driven modelling, and artefact reuse to support the semi-automated generation of verifiable formal specifications.

\noindent\textbf{State-of-the-Art Synthesis:} Through a focused literature review, we categorise and compare (Section 2, supported by Appendix A) recent LLM-based tools and techniques—such as Req2Spec, SpecGen, AssertLLM, and nl2spec—highlighting their approaches, strengths, and limitations in requirement formalisation.

\noindent\textbf{Experimental Evaluation:} We include empirical evaluations (Appendices B and C) comparing multiple SMT solvers (Alt-Ergo, Z3, CVC4, CVC5) in terms of specification verification success and execution time, using the Frama-C PathCrawler tool and standard program sets.

\noindent\textbf{Highlighting Gaps and Future Directions:} Based on our synthesis, we outline critical open problems such as prompt instability, fragility of formal outputs, and the need for domain-specific context grounding (Appendix D). These pave the way for developing more robust LLM-based formal specification pipelines.

\noindent\textbf{Positioning for Long-Term Vision:} As a position paper, our work serves as a foundational step toward a longer-term vision of trustworthy, LLM-assisted formal methods tooling that bridges the expertise gap in safety-critical software development.
 
The structure of the paper is: Section \ref{sec:soa} describes a focused state of the art. Section \ref{sec:challenges} outlines the challenges and future directions for the research, supported by Appendix \ref{app:initial_setup} that outlines the experiment setup and analysis performed on an example of the PathCrawler tool of Frama-C where we have re-simulated the methodology presented in \cite{Granberry2025a}. This analysis is based on four provers available in  Frama-C i.e. Alt-Ergo, Z3, CVC4 and CVC5. Appendix \ref{app:exectime_analysis} presents the execution time comparison for these provers on the programs provided in \cite{FramaCISOLA2024}. Section \ref{sec:discussion} discusses our plans for future work and Section \ref{sec:conclusions} concludes the paper.

\section{State of the Art}
\label{sec:soa}

This section synthesises what we found to be the state of the art. The main research questions for conducting our systematic literature review on the topic were as follows:

\noindent\textbf{RQ1:} What methodologies leverage Large Language Models (LLMs) to transform natural language software requirements into formal notations?

\noindent\textbf{RQ2:} What are the emerging trends and future research directions in using LLMs for formal requirements formalisation?  

Here, we summarise our main findings. For a comprehensive overview of the literature survey, including detailed comparisons and categorised insights, we encourage readers to consult the accompanying GitHub repository, which presents the full set of supporting tables.  

GPT-3.5 has assisted requirement analysis for code verification \cite{10500073}, while Explanation-Refiner integrates LLMs with theorem provers for NLI validation and iterative correction \cite{quan2024verificationrefinementnaturallanguage}. Evaluation of GPT-4o with VeriFast shows generation of functional specifications, though verification remains limited due to redundancy and failed assertions \cite{fan2025evaluatingabilitylargelanguage}.

The nl2spec tool supports interactive synthesis from unstructured requirements \cite{cosler2023nl2specinteractivelytranslatingunstructured}, while the SpecSyn tool improves sequence-to-sequence contract generation with a 21\% accuracy gain \cite{mandal2023largelanguagemodelsbased}. Req2Spec converts 71\% of BOSCH automotive requirements into formal specs \cite{Req2SpecPaper}, while SpecGen uses prompt mutation and verification feedback to improve LLM-generated specifications, succeeding on 279 out of 384 benchmark programs \cite{Ma2024}. In the hardware domain, AssertLLM synthesizes assertions with 89\% correctness via multi-phase prompting and validation \cite{10691792}, while VLSI applications leverage LLMs for spec review and generation in SpecLLM \cite{li2024specllmexploringgenerationreview}. Smart grid requirements have been formalised using GPT-4o and Claude 3.5, achieving F1 scores between 79\% and 94\% \cite{reinpold2024exploringllmsverifyingtechnical}. The trend in F1-scores observed by authors of \cite{reinpold2024exploringllmsverifyingtechnical} suggested that GPT-4o and Claude 3.5 not only maintain robustness but may actually benefit from increased system specification complexity, highlighting a potential alignment between model reasoning depth and specification richness. This aspect not mirrored in Gemini 1.5 or GPT-3.5-turbo and warranting further investigation. NASA's software verification effort surfaced requirement errors, demonstrating practical utility of LLM-in-the-loop workflows \cite{10.1002/spe.430}.

NL-to-LTL translation has seen progress via few-shot prompting and dynamic reasoning \cite{10684640}, achieving 94.4\% accuracy. Likewise, NL-to-JML contract synthesis for Java programs has been explored with promising results \cite{10207159}. Historical systems like RSL \cite{nowakowski2013requirements}, ARSENAL \cite{Ghosh2016}, and RML \cite{Greenspan1986} demonstrated early rule-based and logic-based extraction pipelines, while hybrid neuro-symbolic systems offer greater reliability. SAT-LLM couples SMT solvers with LLMs to detect inconsistencies with F1 of 0.91 \cite{10.1145/3691620.3695302} and LeanDojo, ReProver, and Thor enhance formal proving via retrieval-augmented generation and LLM-guided reasoning \cite{Yang2023,Jiang2022}. IDE-integrated efforts like those combining Copilot with PathCrawler and EVA demonstrate semi-automated ACSL specification generation \cite{Granberry2025,Granberry2025a}.

As we expected, assertion-level synthesis shows better reliability than full contract generation. For example, Laurel generates assertions for Dafny with over 50\% success \cite{mugnier2024laurelgeneratingdafnyassertions}, and AssertLLM exceeds 89\% correctness when guided by contract type and context. Full specifications are more error-prone, often requiring multiple prompt iterations or external validation \cite{Ma2024}.

LLM selection and prompting strategies critically affect performance. While zero-shot prompting is strong in base performance \cite{Kojima2022}, one-shot \cite{li2024oneshotlearninginstructiondata} and few-shot \cite{zhang2023selfconvincedpromptingfewshotquestion} offer alternative trade-offs. Chain-of-Thought (CoT) prompting improves logical flow via intermediate steps \cite{Jason2022}, and like Structured CoT (SCoT) \cite{10.1145/3690635} it can suffer from context decay in long prompts (“lost-in-the-middle”) \cite{Chen2024}, yet zero-shot often remains competitive \cite{NEURIPS2022c4025018}.

Advanced prompting methods like Automate-CoT generate CoT examples automatically \cite{DBLP:conf/emnlp/ShumDZ23}, while Reprompting uses Gibbs sampling to escape prompt local optima \cite{DBLP:journals/corr/abs-2305-09993}. Structured prompting with graphs and trees improves reasoning robustness and efficiency \cite{Besta2024}. RAG (Retrieval-Augmented Generation) improves grounding for knowledge-intensive synthesis \cite{Lewis2020}. \cite{Sahoo2025} discusses a wider range of almost 30 prompting strategies, some of which seem not to have been explored in relation to formalising specifications. Of course, this does not include related approaches such as fine-tuning LLM via LoRA adaptation training \cite{Hu2022}, but this may only be applicable when there is access to the LLM's architecture and weights. Additionally, reinforcement learning (RL) may help with specific challenges, opening even more avenues for exploration. 

\noindent\textbf{Key Observations:}

Based on our literature survey, we proceed with some key observations. 
We observed a significant difference between the success rates of assertion generation and full contract synthesis using LLMs. AssertLLM \cite{10691792} and Laurel \cite{mugnier2024laurelgeneratingdafnyassertions} achieved high accuracy in generating helper assertions for programs written in Dafny language \cite{leino2010dafny} and design-specific verification statements. These tools achieved an accuracy of 89\% and over 50\%, operating at a local level on source code or isolated signals. On the other hand, \cite{10207159} reported that while generating formal specifications for Java Modelling Language (JML) contracts or temporal logic formulas ended up in frequent verification failures by the SMT solvers embedded in OpenJML \cite{cok2011openjml}. This happened even if the output appeared semantically sound, leading to the conclusion of disparity between human-readable correctness and automated formal verification, especially when the source code was written for complex tasks. 

In general, we observed that tasks with small scope and well-defined semantics yield better results where the limited context in these assertions helps in improving verification accuracy \cite{Ma2024, mandal2023largelanguagemodelsbased}. LLMs handle such tasks more reliably due to reduced ambiguity and fewer dependencies on broader system knowledge. On the other hand, for larger program segments, end-to-end contract synthesis involves multiple interacting components or function bodies. It demands a deeper understanding of program semantics, logic, and behaviour over time. SpecGen \cite{Ma2024} and SpecSyn \cite{mandal2023largelanguagemodelsbased} presented significant progress in tackling such challenges. However, their outputs require post-processing steps, such as mutation operators or human-in-the-loop (software testing experts were involved), before the generated outputs are usable for formal verification. 

The effort accuracy is influenced by tool design and integration. For example, tools like nl2spec \cite{cosler2023nl2specinteractivelytranslatingunstructured} improve generated specification quality through step-by-step refinement, adopting iterative and user-in-the-loop approach to help address some LLM limitations. Similarly, prompt engineering techniques utilising guided templates or Chain-of-Thought (CoT) \cite{fan2025evaluatingabilitylargelanguage, 10.1145/3643763, 10656469} promised improved output coherence and correctness. These strategies work well in scenarios involving localised tasks, such as assertion synthesis or narrow-scope descriptions. As the program size and complexity of the specification goal increase, the chances of ambiguity, under-specification, and logical inconsistency increase. Therefore, we conclude that the current LLM architectures excel in focused, declarative tasks but require augmentation for broader specification goals. However, \cite{Porshnev2025ImplicitBias} showed that different versions of language models including LLM, can vary greatly in their responses to the same queries, suggesting that much experimental work might be required to achieve optimum results.

We conclude from our synthesis of the current literature that the research trend is increasing in combining the strength of LLMs, symbolic reasoning and iterative user interaction. At the moment, assertion generation dominates in terms of accuracy and usability, but the parallel efforts of better prompt design, domain-specific fine-tuning, and verifier-in-the-loop are closely matching the performance of the process. The challenges of abstraction and consistency drive research efforts in this domain.

\section{Challenges and Future Directions}
\label{sec:challenges}

Based on our finding, we summarise five key challenges that we have identified, as well as our research goals with brief description given in Table \ref{tab:challenges_future}. A detailed description of these is included in Appendix \ref{app:challenges_future}. 

\begin{table}[ht]
\centering
\begin{tabular}{|p{7.5cm}|p{7.5cm}|}
\hline
\textbf{Challenges} & \textbf{Future Directions} \\
\hline
\textbf{C1: Semantic Ambiguity} \newline Ambiguity in natural language due to context-dependence and jargon affects requirement interpretation. Needs structured knowledge and human-in-the-loop refinement. &
\textbf{F1: Human-in-the-loop Formalisation} \newline Combines LLM support with domain expert oversight to improve accuracy, reduce ambiguity, and increase trust via feedback and interactive refinements. \\
\hline
\textbf{C2: Lack of Ground Truth Datasets} \newline Absence of standardised, annotated datasets limits model training, reproducibility, and scalability. &
\textbf{F3: Standardised Benchmarks} \newline Creation of high-quality, domain-diverse datasets will enable consistent evaluation and push the field forward. \\
\hline
\textbf{C3: Tool Interoperability} \newline Formal verification tools lack standard interfaces and integration capabilities, hampering automation. &
\textbf{F4: Neuro-symbolic Reasoning} \newline Combines neural flexibility with symbolic precision to improve integration, consistency, and constraint enforcement. \\
\hline
\textbf{C4: Traceability Across Artefacts} \newline Difficult to maintain consistent trace links between text, models, code, and tests over lifecycle. &
\textbf{F5: Interactive Traceability Tools} \newline Tools that enable visual navigation, version tracking, and LLM-assisted trace linking improve usability and compliance. \\
\hline
\textbf{C5: Explainability and User Trust} \newline Limited transparency in LLM-generated outputs reduces trust, especially in safety-critical domains. &
\textbf{F2: Multi-modal Artefact Alignment} \newline Integrating diverse input types (text, diagrams, tables) through semantic matching increases clarity and confidence in outputs. \\
\hline
\end{tabular}
\caption{Challenges and Corresponding Future Directions for VERIFYAI Project}
\label{tab:challenges_future}
\end{table}

Semantic ambiguity (C1) due to natural language remains a critical issue, needing structured domain knowledge and improved human-in-the-loop interventions. The lack of publicly available, high-quality datasets (C2) hinders model training, reproducibility, and scalability. Tool interoperability (C3) is impeded by incompatible formats and absence of standardised interfaces, complicating automation. Ensuring traceability across artefact lifecycles (C4) is essential but difficult without explainable and collaborative processes. In addition, explainability and user trust (C5) are limited by opaque model behavior and insufficient rationale in outputs. To address these, human-in-the-loop formalisation (F1) offers controlled semi-automation and improved trust, while multi-modal artefact alignment (F2) enables contextual completeness via diverse input formats. The creation of standardised benchmarks (F3) would mitigate dataset-related limitations and promote progress. Neuro-symbolic reasoning (F4) blends LLM flexibility with logic-based precision, enhancing model reliability. Finally, interactive traceability tools (F5) that support collaboration, visual navigation, and auditability are crucial for regulated and complex software domains. 

\section{Embedding LLMs in the Specification Generation} 
\label{sec:discussion}
\subsection{Planned Evaluation of Prompting Strategies}

A systematic and quantitative evaluation of prompting strategies is the central part of our planned research. In particular, we intend to compare zero-shot, one-shot, few-shot, and Chain-of-Thought (CoT) prompting across both assertion-level and full-contract generation tasks. Prior work in the literature [26–35] suggests that prompt type and  articulation can significantly influence specification quality. However, a comprehensive evaluation requires a larger and more diverse benchmark than we currently report.

Future experiments will therefore assess prompt sensitivity using precision, recall, and F1-based correctness metrics, and investigate robustness under small variations of prompt formulation. We anticipate that assertion-level tasks will prove more stable under prompt rephrasing, whereas full-contract synthesis may show higher variability — an observation that motivates deeper analysis in follow-up work. Our current contribution is to highlight the importance of prompt design in our work and to outline how this dimension will be systematically investigated going forward.

\subsection{Human-in-the-Loop Integration and Interoperability Considerations}

Our experiments currently implement the early stages of a human-in-the-loop process through manual review. In both the Tritype baseline and PathCrawler-augmented workflows, every LLM-generated ACSL specification was checked by at least one author with formal methods expertise. This review ensured (i) semantic alignment with intended behavior, (ii) logical completeness, and (iii) iterative refinement by feeding corrected fragments back into the prompts. Although active learning is not yet integrated, our project approach anticipates it: revised specifications, along with their source code and verification results, can be versioned and selectively reused for retraining or fine-tuning.

\begin{figure}[ht]
    \centering
    \begin{subfigure}[t]{0.48\textwidth}
        \centering
        \begin{tikzpicture}[scale=0.85,
            every node/.style={font=\normalsize, align=center, draw=black, fill=gray!40},
            node distance=1.2cm]
            
            \node (start) [rectangle, rounded corners] {Input C Program};
            \node (pathcrawler) [below of=start, rectangle] {PathCrawler Analysis};
            \node (outputs) [below of=pathcrawler, rectangle] {Symbolic Paths + I/O Examples};
            \node (llm) [below of=outputs, rectangle] {Prompt LLM for ACSL Specs};
            \node (annotated) [below of=llm, rectangle] {Annotated C Code (with ACSL)};
            \node (verify) [below of=annotated, rectangle] {Frama-C WP + SMT Solvers};
            \node (results) [below of=verify, rectangle] {Verification Goals + Outcomes};

            \draw[->, very thick] (start) -- (pathcrawler);
            \draw[->, very thick] (pathcrawler) -- (outputs);
            \draw[->, very thick] (outputs) -- (llm);
            \draw[->, very thick] (llm) -- (annotated);
            \draw[->, very thick] (annotated) -- (verify);
            \draw[->, very thick] (verify) -- (results);
        \end{tikzpicture}
        \caption{Methodology of the initial experiments following the approach of \cite{Granberry2025a}, combining LLM with symbolic analysis tools in the Frama-C ecosystem. The workflow integrates path-based I/O examples and verification outputs to guide the generation of context-aware ACSL specifications.}
        \label{fig:methodology_diagram}
    \end{subfigure}%
    \hfill
    \begin{subfigure}[t]{0.48\textwidth}
        \centering
        \begin{tikzpicture}[scale=0.85,
            every node/.style={font=\small, align=center, draw=black, fill=gray!40},
            node distance=1.5cm]
            
            \node (nlreq) [rectangle, rounded corners] {Input:\\Natural Language Requirements\\+ Domain Ontologies};
            \node (prompts) [below of=nlreq, rectangle] {Different Prompt Strategies\\(zero-shot, few-shot, CoT)};
            \node (llm) [below of=prompts, rectangle] {LLM Specification\\Generation};
            \node (intermediate) [below of=llm, rectangle] {Tool-Neutral Intermediate\\Format (JSON-LD)};
            \node (verify) [below of=intermediate, rectangle] {Verification Tools\\+ Symbolic Reasoning Output};
            \node (human) [below of=verify, rectangle] {Human-in-the-Loop Validation\\+ Example Collection};

            \draw[->, very thick] (nlreq) -- (prompts);
            \draw[->, very thick] (prompts) -- (llm);
            \draw[->, very thick] (llm) -- (intermediate);
            \draw[->, very thick] (intermediate) -- (verify);
            \draw[->, very thick] (verify) -- (human);
						
						\draw[->, very thick, bend left=35] (human.west) to (prompts.west);
        \end{tikzpicture}
        \caption{\textbf{Proposed Workflow for the VERIFYAI pipeline:} Natural-language requirements and domain ontologies are combined with prompt strategies to generate formal specifications via LLMs. Outputs in JSON-LD feed verification tools, with symbolic and human feedback refining results.}
        \label{fig:VERIFYAI-pipeline}
    \end{subfigure}
    \caption{Methodology adopted for initial and future experiments}
\end{figure}

Figure~\ref{fig:VERIFYAI-pipeline} shows our proposed workflow based on the state of the art and figured out challenges in Sections~\ref{sec:soa} and~\ref{sec:challenges}. Natural language requirements and domain ontologies form the input, grounding the LLM in the target context. Different prompt strategies (zero-shot, few-shot, and Chain-of-Thought) shape how inputs are presented for specification generation. Outputs are stored in a \textit{tool-neutral intermediate format} (JSON-LD), which can be translated into the syntax required by verification tools. These tools check the generated specifications, while symbolic reasoning provides constraint-based feedback that helps refine them. Human reviewers remain part of the loop, validating results, focusing on manageable units, and collecting useful examples for future adaptation. The design is modular, so new reasoning engines, prompt methods, or domain adapters can be added without changing the overall pipeline.

For interoperability, the pipeline currently uses custom scripts to convert LLM outputs to tool-specific formats (e.g., ACSL for Frama-C, JML for OpenJML). While functional, this approach is not generalisable. As part of VERIFYAI, we plan to design a lightweight JSON-LD-based schema that maps to multiple formal languages. This would allow LLM outputs to be stored in a tool-agnostic format and exported to different targets, potentially supporting Frama-C, OpenJML, Dafny, and others with minimal per-tool changes.

\subsection{Symbolic Reasoning, Traceability, and Scalability Outlook}

In our initial Tritype experiments, symbolic solvers were used only for post-generation verification. We see potential for tighter integration, where solver feedback—such as unsatisfiable path conditions—could guide the LLM during generation, reducing logical errors before verification. Currently, traceability relies on manual annotations linking specification fragments to source code and natural-language requirements. We are exploring semi-automated approaches where the LLM proposes initial links for human validation. These links could be stored in a graph database to support version-aware navigation and visual trace maps, which would be particularly valuable in regulated settings.

We aim to build datasets that combine real-world, diverse requirements with verified specifications and execution traces. Currently, we are curating a small seed set from open-source safety-critical software, supplemented with synthetic examples to cover edge cases. While synthetic data is useful, it lacks the nuances of industrial requirements, so mixed datasets appear most promising. VERIFYAI’s prototype handles single-module programs well, but multi-module systems pose memory and latency challenges. To manage complexity, we plan to employ hierarchical specification synthesis, verifying each module independently prior to integration. Additionally, we aim to release an annotated subset of the dataset to support transparency and reproducibility.


\section{Conclusions}
\label{sec:conclusions}
This paper identifies clear and concise challenges and prospects in leveraging LLMs for formal software requirements, based on our initial anaysis of the literature. 
Semantic ambiguity, lack of ground truth data, tool interoperability, lifecycle traceability, and explainability all present significant barriers to full automation. However, each challenge also points to fertile ground for innovation. Future directions such as human-in-the-loop systems, multi-modal alignment, standardised benchmarks, neuro-symbolic reasoning, and interactive traceability tools offer practical and scalable paths forward. As a final remark, we can say that as AI and formal methods continue to converge, interdisciplinary collaboration will be key to bridging the gaps and turning conceptual advances into robust, real-world solutions.

\section{Acknowledgements}
This work is partly funded by the ADAPT Research Centre for AI-Driven Digital Content Technology, which is funded by Research Ireland through the Research Ireland Centres Programme and is co funded under the European Regional Development Fund (ERDF) through Grant 13/RC/2106 P2. The submission aligns with Digital Content Transformation (DCT) thread of the ADAPT research centre.
\balance

\bibliographystyle{plain}
\bibliography{overlayConfBib2}

\begin{thebibliography}{47}
\expandafter\ifx\csname natexlab\endcsname\relax\def\natexlab#1{#1}\fi
\providecommand{\url}[1]{\texttt{#1}}
\providecommand{\href}[2]{#2}
\providecommand{\path}[1]{#1}
\providecommand{\DOIprefix}{doi:}
\providecommand{\ArXivprefix}{arXiv:}
\providecommand{\URLprefix}{URL: }
\providecommand{\Pubmedprefix}{pmid:}
\providecommand{\doi}[1]{\href{http://dx.doi.org/#1}{\path{#1}}}
\providecommand{\Pubmed}[1]{\href{pmid:#1}{\path{#1}}}
\providecommand{\bibinfo}[2]{#2}
\ifx\xfnm\relax \def\xfnm[#1]{\unskip,\space#1}\fi
\bibitem[{Huisman et~al.(2024)Huisman, Gurov, and Malkis}]{Huisman2024}
\bibinfo{author}{M.~Huisman}, \bibinfo{author}{D.~Gurov},
  \bibinfo{author}{A.~Malkis}, \bibinfo{title}{Formal methods: From academia to
  industrial practice. a travel guide}, \bibinfo{year}{2024}. \URLprefix
  \url{https://arxiv.org/abs/2002.07279}.
  \href{http://arxiv.org/abs/2002.07279}{{\tt arXiv:2002.07279}}.
\bibitem[{Beg et~al.(2025)Beg, O'Donoghue, and Monahan}]{Beg2025}
\bibinfo{author}{A.~Beg}, \bibinfo{author}{D.~O'Donoghue},
  \bibinfo{author}{R.~Monahan},
\newblock \bibinfo{title}{Formalising software requirements using large
  language models}  (\bibinfo{year}{2025}). \URLprefix
  \url{https://arxiv.org/abs/2506.10704}.
  \href{http://arxiv.org/abs/2506.10704}{{\tt arXiv:2506.10704}}.
\bibitem[{Granberry et~al.(2025)Granberry, Ahrendt, and
  Johansson}]{Granberry2025a}
\bibinfo{author}{G.~Granberry}, \bibinfo{author}{W.~Ahrendt},
  \bibinfo{author}{M.~Johansson},
\newblock \bibinfo{title}{Specify what? enhancing neural specification
  synthesis by symbolic methods},
\newblock in: \bibinfo{editor}{N.~Kosmatov}, \bibinfo{editor}{L.~Kov{\'a}cs}
  (Eds.), \bibinfo{booktitle}{Integrated Formal Methods},
  \bibinfo{publisher}{Springer Nature Switzerland}, \bibinfo{address}{Cham},
  \bibinfo{year}{2025}, pp. \bibinfo{pages}{307--325}.
\bibitem[{Robles et~al.(2024)Robles, Kosmatov, Prevosto, and
  Le~Gall}]{FramaCISOLA2024}
\bibinfo{author}{V.~Robles}, \bibinfo{author}{N.~Kosmatov},
  \bibinfo{author}{V.~Prevosto}, \bibinfo{author}{P.~Le~Gall},
\newblock \bibinfo{title}{High-level program properties in frama-c: Definition,
  verification and deduction},
\newblock in: \bibinfo{booktitle}{Leveraging Applications of Formal Methods,
  Verification and Validation. Specification and Verification: 12th
  International Symposium, ISoLA 2024, Crete, Greece, October 27–31, 2024,
  Proceedings, Part III}, \bibinfo{publisher}{Springer-Verlag},
  \bibinfo{address}{Berlin, Heidelberg}, \bibinfo{year}{2024}, p.
  \bibinfo{pages}{159–177}. \URLprefix
  \url{https://doi.org/10.1007/978-3-031-75380-0_10}.
  \DOIprefix\doi{10.1007/978-3-031-75380-0_10}.
\bibitem[{Couder et~al.(2024)Couder, Gomez, and Ochoa}]{10500073}
\bibinfo{author}{J.~O. Couder}, \bibinfo{author}{D.~Gomez},
  \bibinfo{author}{O.~Ochoa},
\newblock \bibinfo{title}{Requirements verification through the analysis of
  source code by large language models},
\newblock in: \bibinfo{booktitle}{SoutheastCon 2024}, \bibinfo{year}{2024}, pp.
  \bibinfo{pages}{75--80}.
  \DOIprefix\doi{10.1109/SoutheastCon52093.2024.10500073}.
\bibitem[{Quan et~al.(2024)Quan, Valentino, Dennis, and
  Freitas}]{quan2024verificationrefinementnaturallanguage}
\bibinfo{author}{X.~Quan}, \bibinfo{author}{M.~Valentino},
  \bibinfo{author}{L.~A. Dennis}, \bibinfo{author}{A.~Freitas},
  \bibinfo{title}{Verification and refinement of natural language explanations
  through llm-symbolic theorem proving}, \bibinfo{year}{2024}. \URLprefix
  \url{https://arxiv.org/abs/2405.01379}.
  \href{http://arxiv.org/abs/2405.01379}{{\tt arXiv:2405.01379}}.
\bibitem[{Fan et~al.(2025)Fan, Rego, Hu, Dod, Ni, Xie, DiVincenzo, and
  Tan}]{fan2025evaluatingabilitylargelanguage}
\bibinfo{author}{W.~Fan}, \bibinfo{author}{M.~Rego}, \bibinfo{author}{X.~Hu},
  \bibinfo{author}{S.~Dod}, \bibinfo{author}{Z.~Ni}, \bibinfo{author}{D.~Xie},
  \bibinfo{author}{J.~DiVincenzo}, \bibinfo{author}{L.~Tan},
  \bibinfo{title}{Evaluating the ability of large language models to generate
  verifiable specifications in verifast}, \bibinfo{year}{2025}. \URLprefix
  \url{https://arxiv.org/abs/2411.02318}.
  \href{http://arxiv.org/abs/2411.02318}{{\tt arXiv:2411.02318}}.
\bibitem[{Cosler et~al.(2023)Cosler, Hahn, Mendoza, Schmitt, and
  Trippel}]{cosler2023nl2specinteractivelytranslatingunstructured}
\bibinfo{author}{M.~Cosler}, \bibinfo{author}{C.~Hahn},
  \bibinfo{author}{D.~Mendoza}, \bibinfo{author}{F.~Schmitt},
  \bibinfo{author}{C.~Trippel}, \bibinfo{title}{nl2spec: Interactively
  translating unstructured natural language to temporal logics with large
  language models}, \bibinfo{year}{2023}. \URLprefix
  \url{https://arxiv.org/abs/2303.04864}.
  \href{http://arxiv.org/abs/2303.04864}{{\tt arXiv:2303.04864}}.
\bibitem[{Mandal et~al.(2023)Mandal, Chethan, Janfaza, Mahmud, Anderson, Turek,
  Tithi, and Muzahid}]{mandal2023largelanguagemodelsbased}
\bibinfo{author}{S.~Mandal}, \bibinfo{author}{A.~Chethan},
  \bibinfo{author}{V.~Janfaza}, \bibinfo{author}{S.~M.~F. Mahmud},
  \bibinfo{author}{T.~A. Anderson}, \bibinfo{author}{J.~Turek},
  \bibinfo{author}{J.~J. Tithi}, \bibinfo{author}{A.~Muzahid},
  \bibinfo{title}{Large language models based automatic synthesis of software
  specifications}, \bibinfo{year}{2023}. \URLprefix
  \url{https://arxiv.org/abs/2304.09181}.
  \href{http://arxiv.org/abs/2304.09181}{{\tt arXiv:2304.09181}}.
\bibitem[{Nayak et~al.(2022)Nayak, Timmapathini, Murali, Ponnalagu, Venkoparao,
  and Post}]{Req2SpecPaper}
\bibinfo{author}{A.~Nayak}, \bibinfo{author}{H.~P. Timmapathini},
  \bibinfo{author}{V.~Murali}, \bibinfo{author}{K.~Ponnalagu},
  \bibinfo{author}{V.~G. Venkoparao}, \bibinfo{author}{A.~Post},
\newblock \bibinfo{title}{Req2spec: Transforming software requirements into
  formal specifications using natural language processing},
\newblock in: \bibinfo{booktitle}{Requirements Engineering: Foundation for
  Software Quality: 28th International Working Conference, REFSQ 2022,
  Birmingham, UK, March 21–24, 2022, Proceedings},
  \bibinfo{publisher}{Springer-Verlag}, \bibinfo{address}{Berlin, Heidelberg},
  \bibinfo{year}{2022}, p. \bibinfo{pages}{87–95}.
\bibitem[{Ma et~al.(2024)Ma, Liu, Li, Xie, and Bu}]{Ma2024}
\bibinfo{author}{L.~Ma}, \bibinfo{author}{S.~Liu}, \bibinfo{author}{Y.~Li},
  \bibinfo{author}{X.~Xie}, \bibinfo{author}{L.~Bu},
\newblock \bibinfo{title}{Specgen: Automated generation of formal program
  specifications via large language models}  (\bibinfo{year}{2024}). \URLprefix
  \url{https://arxiv.org/abs/2401.08807}.
  \href{http://arxiv.org/abs/2401.08807}{{\tt arXiv:2401.08807}}.
\bibitem[{Fang et~al.(2024)Fang, Li, Li, Yan, Liu, Zhang, and Xie}]{10691792}
\bibinfo{author}{W.~Fang}, \bibinfo{author}{M.~Li}, \bibinfo{author}{M.~Li},
  \bibinfo{author}{Z.~Yan}, \bibinfo{author}{S.~Liu},
  \bibinfo{author}{H.~Zhang}, \bibinfo{author}{Z.~Xie},
\newblock \bibinfo{title}{Assertllm: Generating hardware verification
  assertions from design specifications via multi-llms},
\newblock in: \bibinfo{booktitle}{2024 IEEE LLM Aided Design Workshop (LAD)},
  \bibinfo{year}{2024}, pp. \bibinfo{pages}{1--1}.
  \DOIprefix\doi{10.1109/LAD62341.2024.10691792}.
\bibitem[{Li et~al.(2024)Li, Fang, Zhang, and
  Xie}]{li2024specllmexploringgenerationreview}
\bibinfo{author}{M.~Li}, \bibinfo{author}{W.~Fang}, \bibinfo{author}{Q.~Zhang},
  \bibinfo{author}{Z.~Xie}, \bibinfo{title}{Specllm: Exploring generation and
  review of vlsi design specification with large language model},
  \bibinfo{year}{2024}. \URLprefix \url{https://arxiv.org/abs/2401.13266}.
  \href{http://arxiv.org/abs/2401.13266}{{\tt arXiv:2401.13266}}.
\bibitem[{Reinpold et~al.(2024)Reinpold, Schieseck, Wagner, Gehlhoff, and
  Fay}]{reinpold2024exploringllmsverifyingtechnical}
\bibinfo{author}{L.~M. Reinpold}, \bibinfo{author}{M.~Schieseck},
  \bibinfo{author}{L.~P. Wagner}, \bibinfo{author}{F.~Gehlhoff},
  \bibinfo{author}{A.~Fay}, \bibinfo{title}{Exploring llms for verifying
  technical system specifications against requirements}, \bibinfo{year}{2024}.
  \URLprefix \url{https://arxiv.org/abs/2411.11582}.
  \href{http://arxiv.org/abs/2411.11582}{{\tt arXiv:2411.11582}}.
\bibitem[{Gervasi and Nuseibeh(2002)}]{10.1002/spe.430}
\bibinfo{author}{V.~Gervasi}, \bibinfo{author}{B.~Nuseibeh},
\newblock \bibinfo{title}{Lightweight validation of natural language
  requirements},
\newblock \bibinfo{journal}{Softw. Pract. Exper.} \bibinfo{volume}{32}
  (\bibinfo{year}{2002}) \bibinfo{pages}{113–133}. \URLprefix
  \url{https://doi.org/10.1002/spe.430}. \DOIprefix\doi{10.1002/spe.430}.
\bibitem[{Xu et~al.(2024)Xu, Feng, and Miao}]{10684640}
\bibinfo{author}{Y.~Xu}, \bibinfo{author}{J.~Feng}, \bibinfo{author}{W.~Miao},
\newblock \bibinfo{title}{Learning from failures: Translation of natural
  language requirements into linear temporal logic with large language models},
\newblock in: \bibinfo{booktitle}{2024 IEEE 24th International Conference on
  Software Quality, Reliability and Security (QRS)}, \bibinfo{year}{2024}, pp.
  \bibinfo{pages}{204--215}. \DOIprefix\doi{10.1109/QRS62785.2024.00029}.
\bibitem[{Leong and Barbosa(2023)}]{10207159}
\bibinfo{author}{I.~T. Leong}, \bibinfo{author}{R.~Barbosa},
\newblock \bibinfo{title}{Translating natural language requirements to formal
  specifications: A study on gpt and symbolic nlp},
\newblock in: \bibinfo{booktitle}{2023 53rd Annual IEEE/IFIP International
  Conference on Dependable Systems and Networks Workshops (DSN-W)},
  \bibinfo{year}{2023}, pp. \bibinfo{pages}{259--262}.
  \DOIprefix\doi{10.1109/DSN-W58399.2023.00065}.
\bibitem[{Nowakowski et~al.(2013)Nowakowski, Śmiałek, Ambroziewicz, and
  Straszak}]{nowakowski2013requirements}
\bibinfo{author}{W.~Nowakowski}, \bibinfo{author}{M.~Śmiałek},
  \bibinfo{author}{A.~Ambroziewicz}, \bibinfo{author}{T.~Straszak},
\newblock \bibinfo{title}{Requirements-level language and tools for capturing
  software system essence},
\newblock \bibinfo{journal}{Computer Science and Information Systems}
  \bibinfo{volume}{10} (\bibinfo{year}{2013}) \bibinfo{pages}{1499--1524}.
\bibitem[{Ghosh et~al.(2016)Ghosh, Elenius, Li, Lincoln, Shankar, and
  Steiner}]{Ghosh2016}
\bibinfo{author}{S.~Ghosh}, \bibinfo{author}{D.~Elenius},
  \bibinfo{author}{W.~Li}, \bibinfo{author}{P.~Lincoln},
  \bibinfo{author}{N.~Shankar}, \bibinfo{author}{W.~Steiner},
\newblock \bibinfo{title}{Arsenal: Automatic requirements specification
  extraction from natural language},
\newblock in: \bibinfo{editor}{S.~Rayadurgam}, \bibinfo{editor}{O.~Tkachuk}
  (Eds.), \bibinfo{booktitle}{NASA Formal Methods},
  \bibinfo{publisher}{Springer International Publishing},
  \bibinfo{address}{Cham}, \bibinfo{year}{2016}, pp. \bibinfo{pages}{41--46}.
\bibitem[{Greenspan et~al.(1986)Greenspan, Borgida, and
  Mylopoulos}]{Greenspan1986}
\bibinfo{author}{S.~J. Greenspan}, \bibinfo{author}{A.~Borgida},
  \bibinfo{author}{J.~Mylopoulos},
\newblock \bibinfo{title}{A requirements modeling language and its logic},
\newblock \bibinfo{journal}{Information Systems} \bibinfo{volume}{11}
  (\bibinfo{year}{1986}) \bibinfo{pages}{9--23}. \URLprefix
  \url{https://www.sciencedirect.com/science/article/pii/0306437986900207}.
  \DOIprefix\doi{https://doi.org/10.1016/0306-4379(86)90020-7}.
\bibitem[{Fazelnia et~al.(2024)Fazelnia, Mirakhorli, and
  Bagheri}]{10.1145/3691620.3695302}
\bibinfo{author}{M.~Fazelnia}, \bibinfo{author}{M.~Mirakhorli},
  \bibinfo{author}{H.~Bagheri},
\newblock \bibinfo{title}{Translation titans, reasoning challenges:
  Satisfiability-aided language models for detecting conflicting requirements},
\newblock in: \bibinfo{booktitle}{Proceedings of the 39th IEEE/ACM
  International Conference on Automated Software Engineering}, ASE '24,
  \bibinfo{publisher}{Association for Computing Machinery},
  \bibinfo{address}{New York, NY, USA}, \bibinfo{year}{2024}, p.
  \bibinfo{pages}{2294–2298}. \URLprefix
  \url{https://doi.org/10.1145/3691620.3695302}.
  \DOIprefix\doi{10.1145/3691620.3695302}.
\bibitem[{Yang et~al.(2023)Yang, Swope, Gu, Chalamala, Song, Yu, Godil,
  Prenger, and Anandkumar}]{Yang2023}
\bibinfo{author}{K.~Yang}, \bibinfo{author}{A.~Swope}, \bibinfo{author}{A.~Gu},
  \bibinfo{author}{R.~Chalamala}, \bibinfo{author}{P.~Song},
  \bibinfo{author}{S.~Yu}, \bibinfo{author}{S.~Godil}, \bibinfo{author}{R.~J.
  Prenger}, \bibinfo{author}{A.~Anandkumar},
\newblock \bibinfo{title}{Leandojo: Theorem proving with retrieval-augmented
  language models},
\newblock in: \bibinfo{editor}{A.~Oh}, \bibinfo{editor}{T.~Naumann},
  \bibinfo{editor}{A.~Globerson}, \bibinfo{editor}{K.~Saenko},
  \bibinfo{editor}{M.~Hardt}, \bibinfo{editor}{S.~Levine} (Eds.),
  \bibinfo{booktitle}{Advances in Neural Information Processing Systems},
  volume~\bibinfo{volume}{36}, \bibinfo{publisher}{Curran Associates, Inc.},
  \bibinfo{year}{2023}, pp. \bibinfo{pages}{21573--21612}.
\bibitem[{Jiang et~al.(2022)Jiang, Li, Tworkowski, Czechowski,
  Odrzyg\'{o}\'{z}d\'{z}, Mi\l~o\'{s}, Wu, and Jamnik}]{Jiang2022}
\bibinfo{author}{A.~Q. Jiang}, \bibinfo{author}{W.~Li},
  \bibinfo{author}{S.~Tworkowski}, \bibinfo{author}{K.~Czechowski},
  \bibinfo{author}{T.~Odrzyg\'{o}\'{z}d\'{z}},
  \bibinfo{author}{P.~Mi\l~o\'{s}}, \bibinfo{author}{Y.~Wu},
  \bibinfo{author}{M.~Jamnik},
\newblock \bibinfo{title}{Thor: Wielding hammers to integrate language models
  and automated theorem provers},
\newblock in: \bibinfo{editor}{S.~Koyejo}, \bibinfo{editor}{S.~Mohamed},
  \bibinfo{editor}{A.~Agarwal}, \bibinfo{editor}{D.~Belgrave},
  \bibinfo{editor}{K.~Cho}, \bibinfo{editor}{A.~Oh} (Eds.),
  \bibinfo{booktitle}{Advances in Neural Information Processing Systems},
  volume~\bibinfo{volume}{35}, \bibinfo{publisher}{Curran Associates, Inc.},
  \bibinfo{year}{2022}, pp. \bibinfo{pages}{8360--8373}.
\bibitem[{Granberry et~al.(2025)Granberry, Ahrendt, and
  Johansson}]{Granberry2025}
\bibinfo{author}{G.~Granberry}, \bibinfo{author}{W.~Ahrendt},
  \bibinfo{author}{M.~Johansson},
\newblock \bibinfo{title}{Towards integrating copiloting and formal methods},
\newblock in: \bibinfo{editor}{T.~Margaria}, \bibinfo{editor}{B.~Steffen}
  (Eds.), \bibinfo{booktitle}{Leveraging Applications of Formal Methods,
  Verification and Validation. Specification and Verification},
  \bibinfo{publisher}{Springer Nature Switzerland}, \bibinfo{address}{Cham},
  \bibinfo{year}{2025}, pp. \bibinfo{pages}{144--158}.
\bibitem[{Mugnier et~al.(2024)Mugnier, Gonzalez, Jhala, Polikarpova, and
  Zhou}]{mugnier2024laurelgeneratingdafnyassertions}
\bibinfo{author}{E.~Mugnier}, \bibinfo{author}{E.~A. Gonzalez},
  \bibinfo{author}{R.~Jhala}, \bibinfo{author}{N.~Polikarpova},
  \bibinfo{author}{Y.~Zhou}, \bibinfo{title}{Laurel: Generating dafny
  assertions using large language models}, \bibinfo{year}{2024}. \URLprefix
  \url{https://arxiv.org/abs/2405.16792}.
  \href{http://arxiv.org/abs/2405.16792}{{\tt arXiv:2405.16792}}.
\bibitem[{Kojima et~al.(2022)Kojima, Gu, Reid, Matsuo, and
  Iwasawa}]{Kojima2022}
\bibinfo{author}{T.~Kojima}, \bibinfo{author}{S.~S. Gu},
  \bibinfo{author}{M.~Reid}, \bibinfo{author}{Y.~Matsuo},
  \bibinfo{author}{Y.~Iwasawa},
\newblock \bibinfo{title}{Large language models are zero-shot reasoners},
\newblock in: \bibinfo{editor}{S.~Koyejo}, \bibinfo{editor}{S.~Mohamed},
  \bibinfo{editor}{A.~Agarwal}, \bibinfo{editor}{D.~Belgrave},
  \bibinfo{editor}{K.~Cho}, \bibinfo{editor}{A.~Oh} (Eds.),
  \bibinfo{booktitle}{Advances in Neural Information Processing Systems},
  volume~\bibinfo{volume}{35}, \bibinfo{publisher}{Curran Associates, Inc.},
  \bibinfo{year}{2022}, pp. \bibinfo{pages}{22199--22213}.
\bibitem[{Li et~al.(2024)Li, Hui, Xia, Yang, Yang, Zhang, Si, Chen, Liu, Liu,
  Huang, and Li}]{li2024oneshotlearninginstructiondata}
\bibinfo{author}{Y.~Li}, \bibinfo{author}{B.~Hui}, \bibinfo{author}{X.~Xia},
  \bibinfo{author}{J.~Yang}, \bibinfo{author}{M.~Yang},
  \bibinfo{author}{L.~Zhang}, \bibinfo{author}{S.~Si}, \bibinfo{author}{L.-H.
  Chen}, \bibinfo{author}{J.~Liu}, \bibinfo{author}{T.~Liu},
  \bibinfo{author}{F.~Huang}, \bibinfo{author}{Y.~Li},
\newblock \bibinfo{title}{One-shot learning as instruction data prospector for
  large language models}  (\bibinfo{year}{2024}). \URLprefix
  \url{https://arxiv.org/abs/2312.10302}.
  \href{http://arxiv.org/abs/2312.10302}{{\tt arXiv:2312.10302}}.
\bibitem[{Zhang et~al.(2023)Zhang, Cai, Zhang, Zhang, Mao, and
  Wu}]{zhang2023selfconvincedpromptingfewshotquestion}
\bibinfo{author}{H.~Zhang}, \bibinfo{author}{M.~Cai},
  \bibinfo{author}{X.~Zhang}, \bibinfo{author}{C.~J. Zhang},
  \bibinfo{author}{R.~Mao}, \bibinfo{author}{K.~Wu},
\newblock \bibinfo{title}{Self-convinced prompting: Few-shot question answering
  with repeated introspection}  (\bibinfo{year}{2023}). \URLprefix
  \url{https://arxiv.org/abs/2310.05035}.
  \href{http://arxiv.org/abs/2310.05035}{{\tt arXiv:2310.05035}}.
\bibitem[{Wei et~al.(2022)Wei, Wang, Schuurmans, Bosma, ichter, Xia, Chi, Le,
  and Zhou}]{Jason2022}
\bibinfo{author}{J.~Wei}, \bibinfo{author}{X.~Wang},
  \bibinfo{author}{D.~Schuurmans}, \bibinfo{author}{M.~Bosma},
  \bibinfo{author}{b.~ichter}, \bibinfo{author}{F.~Xia},
  \bibinfo{author}{E.~Chi}, \bibinfo{author}{Q.~V. Le},
  \bibinfo{author}{D.~Zhou},
\newblock \bibinfo{title}{Chain-of-thought prompting elicits reasoning in large
  language models},
\newblock in: \bibinfo{editor}{S.~Koyejo}, \bibinfo{editor}{S.~Mohamed},
  \bibinfo{editor}{A.~Agarwal}, \bibinfo{editor}{D.~Belgrave},
  \bibinfo{editor}{K.~Cho}, \bibinfo{editor}{A.~Oh} (Eds.),
  \bibinfo{booktitle}{Advances in Neural Information Processing Systems},
  volume~\bibinfo{volume}{35}, \bibinfo{publisher}{Curran Associates, Inc.},
  \bibinfo{year}{2022}, pp. \bibinfo{pages}{24824--24837}.
\bibitem[{Li et~al.(2025)Li, Li, Li, and Jin}]{10.1145/3690635}
\bibinfo{author}{J.~Li}, \bibinfo{author}{G.~Li}, \bibinfo{author}{Y.~Li},
  \bibinfo{author}{Z.~Jin},
\newblock \bibinfo{title}{Structured chain-of-thought prompting for code
  generation},
\newblock \bibinfo{journal}{ACM Trans. Softw. Eng. Methodol.}
  \bibinfo{volume}{34} (\bibinfo{year}{2025}). \URLprefix
  \url{https://doi.org/10.1145/3690635}. \DOIprefix\doi{10.1145/3690635}.
\bibitem[{Hsieh et~al.(2024)Hsieh, Chuang, Li, Wang, Le, Kumar, Glass, Ratner,
  Lee, Krishna, and Pfister}]{Chen2024}
\bibinfo{author}{C.~Hsieh}, \bibinfo{author}{Y.~Chuang},
  \bibinfo{author}{C.~Li}, \bibinfo{author}{Z.~Wang}, \bibinfo{author}{L.~T.
  Le}, \bibinfo{author}{A.~Kumar}, \bibinfo{author}{J.~R. Glass},
  \bibinfo{author}{A.~Ratner}, \bibinfo{author}{C.~Lee},
  \bibinfo{author}{R.~Krishna}, \bibinfo{author}{T.~Pfister},
\newblock \bibinfo{title}{Found in the middle: Calibrating positional attention
  bias improves long context utilization},
\newblock in: \bibinfo{editor}{L.~Ku}, \bibinfo{editor}{A.~Martins},
  \bibinfo{editor}{V.~Srikumar} (Eds.), \bibinfo{booktitle}{Findings of the
  Association for Computational Linguistics, {ACL} 2024, Bangkok, Thailand and
  virtual meeting, August 11-16, 2024}, \bibinfo{publisher}{Association for
  Computational Linguistics}, \bibinfo{year}{2024}, pp.
  \bibinfo{pages}{14982--14995}. \URLprefix
  \url{https://doi.org/10.18653/v1/2024.findings-acl.890}.
  \DOIprefix\doi{10.18653/V1/2024.FINDINGS-ACL.890}.
\bibitem[{Ye and Durrett(2022)}]{NEURIPS2022c4025018}
\bibinfo{author}{X.~Ye}, \bibinfo{author}{G.~Durrett},
\newblock \bibinfo{title}{The unreliability of explanations in few-shot
  prompting for textual reasoning},
\newblock in: \bibinfo{editor}{S.~Koyejo}, \bibinfo{editor}{S.~Mohamed},
  \bibinfo{editor}{A.~Agarwal}, \bibinfo{editor}{D.~Belgrave},
  \bibinfo{editor}{K.~Cho}, \bibinfo{editor}{A.~Oh} (Eds.),
  \bibinfo{booktitle}{Advances in Neural Information Processing Systems},
  volume~\bibinfo{volume}{35}, \bibinfo{publisher}{Curran Associates, Inc.},
  \bibinfo{year}{2022}, pp. \bibinfo{pages}{30378--30392}.
\bibitem[{Shum et~al.(2023)Shum, Diao, and Zhang}]{DBLP:conf/emnlp/ShumDZ23}
\bibinfo{author}{K.~Shum}, \bibinfo{author}{S.~Diao},
  \bibinfo{author}{T.~Zhang},
\newblock \bibinfo{title}{Automatic prompt augmentation and selection with
  chain-of-thought from labeled data},
\newblock in: \bibinfo{editor}{H.~Bouamor}, \bibinfo{editor}{J.~Pino},
  \bibinfo{editor}{K.~Bali} (Eds.), \bibinfo{booktitle}{Findings of the
  Association for Computational Linguistics: {EMNLP} 2023, Singapore, December
  6-10, 2023}, \bibinfo{publisher}{Association for Computational Linguistics},
  \bibinfo{year}{2023}, pp. \bibinfo{pages}{12113--12139}. \URLprefix
  \url{https://doi.org/10.18653/v1/2023.findings-emnlp.811}.
  \DOIprefix\doi{10.18653/V1/2023.FINDINGS-EMNLP.811}.
\bibitem[{Xu et~al.(2023)Xu, Banburski{-}Fahey, and
  Jojic}]{DBLP:journals/corr/abs-2305-09993}
\bibinfo{author}{W.~Xu}, \bibinfo{author}{A.~Banburski{-}Fahey},
  \bibinfo{author}{N.~Jojic},
\newblock \bibinfo{title}{Reprompting: Automated chain-of-thought prompt
  inference through gibbs sampling},
\newblock \bibinfo{journal}{CoRR} \bibinfo{volume}{abs/2305.09993}
  (\bibinfo{year}{2023}). \URLprefix
  \url{https://doi.org/10.48550/arXiv.2305.09993}.
  \DOIprefix\doi{10.48550/ARXIV.2305.09993}.
  \href{http://arxiv.org/abs/2305.09993}{{\tt arXiv:2305.09993}}.
\bibitem[{Besta et~al.(2024)Besta, Memedi, Zhang, Gerstenberger, Blach, Nyczyk,
  Copik, Kwasniewski, M{\"{u}}ller, Gianinazzi, Kubicek, Niewiadomski, Mutlu,
  and Hoefler}]{Besta2024}
\bibinfo{author}{M.~Besta}, \bibinfo{author}{F.~Memedi},
  \bibinfo{author}{Z.~Zhang}, \bibinfo{author}{R.~Gerstenberger},
  \bibinfo{author}{N.~Blach}, \bibinfo{author}{P.~Nyczyk},
  \bibinfo{author}{M.~Copik}, \bibinfo{author}{G.~Kwasniewski},
  \bibinfo{author}{J.~M{\"{u}}ller}, \bibinfo{author}{L.~Gianinazzi},
  \bibinfo{author}{A.~Kubicek}, \bibinfo{author}{H.~Niewiadomski},
  \bibinfo{author}{O.~Mutlu}, \bibinfo{author}{T.~Hoefler},
\newblock \bibinfo{title}{Topologies of reasoning: Demystifying chains, trees,
  and graphs of thoughts},
\newblock \bibinfo{journal}{CoRR} \bibinfo{volume}{abs/2401.14295}
  (\bibinfo{year}{2024}). \URLprefix
  \url{https://doi.org/10.48550/arXiv.2401.14295}.
  \DOIprefix\doi{10.48550/arXiv.2401.14295}.
  \href{http://arxiv.org/abs/2401.14295}{{\tt arXiv:2401.14295}}.
\bibitem[{Lewis et~al.(2020)Lewis, Perez, Piktus, Petroni, Karpukhin, Goyal,
  K{\"{u}}ttler, Lewis, Yih, Rockt{\"{a}}schel, Riedel, and Kiela}]{Lewis2020}
\bibinfo{author}{P.~Lewis}, \bibinfo{author}{E.~Perez},
  \bibinfo{author}{A.~Piktus}, \bibinfo{author}{F.~Petroni},
  \bibinfo{author}{V.~Karpukhin}, \bibinfo{author}{N.~Goyal},
  \bibinfo{author}{H.~K{\"{u}}ttler}, \bibinfo{author}{M.~Lewis},
  \bibinfo{author}{W.~Yih}, \bibinfo{author}{T.~Rockt{\"{a}}schel},
  \bibinfo{author}{S.~Riedel}, \bibinfo{author}{D.~Kiela},
\newblock \bibinfo{title}{Retrieval-augmented generation for
  knowledge-intensive {NLP} tasks},
\newblock in: \bibinfo{editor}{H.~Larochelle}, \bibinfo{editor}{M.~Ranzato},
  \bibinfo{editor}{R.~Hadsell}, \bibinfo{editor}{M.~Balcan},
  \bibinfo{editor}{H.~Lin} (Eds.), \bibinfo{booktitle}{Advances in Neural
  Information Processing Systems 33: Annual Conference on Neural Information
  Processing Systems 2020, NeurIPS 2020, December 6-12, 2020, virtual},
  \bibinfo{year}{2020}. \URLprefix
  \url{https://proceedings.neurips.cc/paper/2020/hash/6b493230205f780e1bc26945df7481e5-Abstract.html}.
\bibitem[{Sahoo et~al.(2025)Sahoo, Singh, Saha, Jain, Mondal, and
  Chadha}]{Sahoo2025}
\bibinfo{author}{P.~Sahoo}, \bibinfo{author}{A.~K. Singh},
  \bibinfo{author}{S.~Saha}, \bibinfo{author}{V.~Jain},
  \bibinfo{author}{S.~Mondal}, \bibinfo{author}{A.~Chadha},
\newblock \bibinfo{title}{A systematic survey of prompt engineering in large
  language models: Techniques and applications}  (\bibinfo{year}{2025}).
  \URLprefix \url{https://arxiv.org/abs/2402.07927}.
  \href{http://arxiv.org/abs/2402.07927}{{\tt arXiv:2402.07927}}.
\bibitem[{Hu et~al.(2022)Hu, Shen, Wallis, Allen{-}Zhu, Li, Wang, Wang, and
  Chen}]{Hu2022}
\bibinfo{author}{E.~J. Hu}, \bibinfo{author}{Y.~Shen},
  \bibinfo{author}{P.~Wallis}, \bibinfo{author}{Z.~Allen{-}Zhu},
  \bibinfo{author}{Y.~Li}, \bibinfo{author}{S.~Wang},
  \bibinfo{author}{L.~Wang}, \bibinfo{author}{W.~Chen},
\newblock \bibinfo{title}{Lora: Low-rank adaptation of large language models},
\newblock in: \bibinfo{booktitle}{The Tenth International Conference on
  Learning Representations, {ICLR} 2022, Virtual Event, April 25-29, 2022},
  \bibinfo{publisher}{OpenReview.net}, \bibinfo{year}{2022}. \URLprefix
  \url{https://openreview.net/forum?id=nZeVKeeFYf9}.
\bibitem[{Leino(2010)}]{leino2010dafny}
\bibinfo{author}{K.~R.~M. Leino},
\newblock \bibinfo{title}{Dafny: An automatic program verifier for functional
  correctness},
\newblock in: \bibinfo{booktitle}{Proceedings of the 16th International
  Conference on Logic for Programming, Artificial Intelligence, and Reasoning
  (LPAR)}, volume \bibinfo{volume}{6355} of \textit{\bibinfo{series}{Lecture
  Notes in Computer Science}}, \bibinfo{publisher}{Springer},
  \bibinfo{year}{2010}, pp. \bibinfo{pages}{348--370}. \URLprefix
  \url{https://doi.org/10.1007/978-3-642-17511-4\_20}.
  \DOIprefix\doi{10.1007/978-3-642-17511-4\_20}.
\bibitem[{Cok(2011)}]{cok2011openjml}
\bibinfo{author}{D.~R. Cok},
\newblock \bibinfo{title}{Openjml: Software verification for java 7 using jml,
  openjdk, and eclipse},
\newblock in: \bibinfo{booktitle}{NASA Formal Methods (NFM 2011)}, volume
  \bibinfo{volume}{6617} of \textit{\bibinfo{series}{Lecture Notes in Computer
  Science}}, \bibinfo{publisher}{Springer}, \bibinfo{year}{2011}, pp.
  \bibinfo{pages}{472--479}. \URLprefix
  \url{https://doi.org/10.1007/978-3-642-20398-5_35}.
  \DOIprefix\doi{10.1007/978-3-642-20398-5\_35}.
\bibitem[{Misu et~al.(2024)Misu, Lopes, Ma, and Noble}]{10.1145/3643763}
\bibinfo{author}{M.~R.~H. Misu}, \bibinfo{author}{C.~V. Lopes},
  \bibinfo{author}{I.~Ma}, \bibinfo{author}{J.~Noble},
\newblock \bibinfo{title}{Towards ai-assisted synthesis of verified dafny
  methods},
\newblock \bibinfo{journal}{Proc. ACM Softw. Eng.} \bibinfo{volume}{1}
  (\bibinfo{year}{2024}). \URLprefix \url{https://doi.org/10.1145/3643763}.
  \DOIprefix\doi{10.1145/3643763}.
\bibitem[{Yao et~al.(2024)Yao, Liu, Dong, Guo, Hu, Keutzer, Du, Zhou, and
  Zhang}]{10656469}
\bibinfo{author}{J.~Yao}, \bibinfo{author}{Y.~Liu}, \bibinfo{author}{Z.~Dong},
  \bibinfo{author}{M.~Guo}, \bibinfo{author}{H.~Hu},
  \bibinfo{author}{K.~Keutzer}, \bibinfo{author}{L.~Du},
  \bibinfo{author}{D.~Zhou}, \bibinfo{author}{S.~Zhang},
\newblock \bibinfo{title}{Promptcot: Align prompt distribution via adapted
  chain-of-thought},
\newblock in: \bibinfo{booktitle}{2024 IEEE/CVF Conference on Computer Vision
  and Pattern Recognition (CVPR)}, \bibinfo{year}{2024}, pp.
  \bibinfo{pages}{7027--7037}. \DOIprefix\doi{10.1109/CVPR52733.2024.00671}.
\bibitem[{Porshnev et~al.(2025)}]{Porshnev2025ImplicitBias}
\bibinfo{author}{A.~Porshnev}, et~al.,
\newblock \bibinfo{title}{Modelling implicit bias in gender–career
  associations: A systematic comparison of language models},
\newblock \bibinfo{journal}{PsyArXiv}  (\bibinfo{year}{2025}).
  \DOIprefix\doi{10.31234/osf.io/p7hvw\_v1}, \bibinfo{note}{preprint, 22 May
  2025}.
\bibitem[{Tahir et~al.(2025)Tahir, Jahankhani, Tasleem, and
  Hassan}]{systems13070567}
\bibinfo{author}{T.~Tahir}, \bibinfo{author}{H.~Jahankhani},
  \bibinfo{author}{K.~Tasleem}, \bibinfo{author}{B.~Hassan},
\newblock \bibinfo{title}{Cross-project multiclass classification of ears-based
  functional requirements utilizing natural language processing, machine
  learning, and deep learning},
\newblock \bibinfo{journal}{Systems} \bibinfo{volume}{13}
  (\bibinfo{year}{2025}). \URLprefix
  \url{https://www.mdpi.com/2079-8954/13/7/567}.
  \DOIprefix\doi{10.3390/systems13070567}.
\bibitem[{Gregory(2020)}]{Gregory2020}
\bibinfo{author}{K.~Gregory},
\newblock \bibinfo{title}{A dataset describing data discovery and reuse
  practices in research},
\newblock \bibinfo{journal}{Scientific Data} \bibinfo{volume}{7}
  (\bibinfo{year}{2020}) \bibinfo{pages}{232}. \URLprefix
  \url{https://doi.org/10.1038/s41597-020-0569-5}.
  \DOIprefix\doi{10.1038/s41597-020-0569-5}.
\bibitem[{Boetticher et~al.(2007)Boetticher, Menzies, and
  Ostrand}]{PROMISE2007DataSet}
\bibinfo{author}{G.~Boetticher}, \bibinfo{author}{T.~Menzies},
  \bibinfo{author}{T.~Ostrand},
\newblock \bibinfo{title}{\{PROMISE\} repository of empirical software
  engineering data}  (\bibinfo{year}{2007}).
\bibitem[{Hayes et~al.(2018)Hayes, Payne, and Dekhtyar}]{RETRONET2018}
\bibinfo{author}{J.~H. Hayes}, \bibinfo{author}{J.~Payne},
  \bibinfo{author}{A.~Dekhtyar},
\newblock \bibinfo{title}{The requirements tracing on target (retro).net
  dataset}  (\bibinfo{year}{2018}). \URLprefix
  \url{https://arxiv.org/abs/1807.11344}.
  \href{http://arxiv.org/abs/1807.11344}{{\tt arXiv:1807.11344}}.

\end{thebibliography}
\clearpage
\appendix

\section{Work in progress and initial experiment setup}
\label{app:initial_setup}
\begin{figure}[ht]
    \centering
		\includegraphics[width=\linewidth]{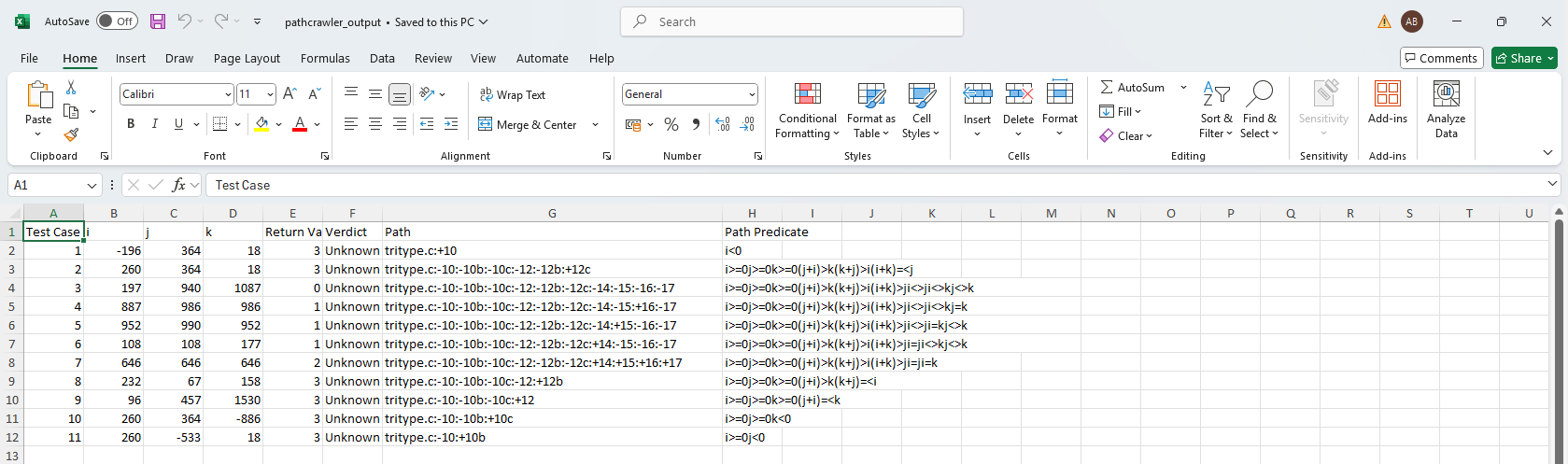}
    \caption{Output of the Pathcrawler for TriType Example}
    \label{fig:pathcrawler_output}
\end{figure}

As initial experiment setup, we re-simulated the methodology presented in \cite{Granberry2025a} by applying it to TriType.c program, which was selected from the online interface of the PathCrawler tool available in the Frama-C ecosystem.  The program was analysed using the original workflow -- combining Large Language Models (LLMs) with symbolic outputs from PathCrawler and EVA -- to generate ACSL specifications. The use of PathCrawler provided concrete path-based input/output examples that guided the LLM toward generating more context-relevant and semantically aligned annotations. Figure \ref{fig:pathcrawler_output} shows the output of the Pathcrawler tool. Our plan is to conduct experiments with simple and complex programs, representing a diverse set of procedural constructs and control flows, allowing us to evaluate the methodology across a realistic and varied set of inputs. A repository of related resources is available at \url{https://github.com/arshadbeg/OVERLAY2025_SupportingDocs.git}. 

\subsection*{Analysis of Tritype Example}

\textbf{Note:} We used OpenAI’s ChatGPT-4o to assist in generating descriptive analysis text based on the output of the PathCrawler tool. All generated content was reviewed and verified by the authors. The function \texttt{Tritype} is:
\begingroup
\tiny
\begin{verbatim}
/* Should return the type of the triangle 
   which has sides of these lengths.
   3 = not a triangle
   2 = equilateral triangle
   1 = isoceles triangle
   0 = other triangle
*/
int Tritype(double i, double j, double k){
  int trityp = 0;
  if (i < 0.0 || j < 0.0 || k < 0.0)          // line 10  
    return 3;
  if (i + j <= k || j + k <= i || k + i <= j) // line 12 
    return 3;    
  if (i == j) trityp = trityp + 1;            // line 14
  if (i == k) trityp = trityp + 1;            // line 15
  if (j == k) trityp = trityp + 1;            // line 16
  if (trityp >= 2)                            // line 17
      trityp = 2;
  return trityp;
}
\end{verbatim}
\endgroup

PathCrawler successfully instrumented and analysed the \texttt{Tritype} function using the all-path coverage criterion. While 100\% branch coverage was achieved, none of the 11 generated test cases were categorised as successful or failed, leaving all as "unknown." This implies that the tool executed all branches in the code, yet could not conclusively determine the verdict for any specific path, possibly due to incomplete post-condition specifications or ambiguity in return value interpretation. The test harness was properly compiled, and the environment was initialised and cleaned up, suggesting that the tool functioned without interruptions. However, the presence of infeasible paths (8 of 19) emphasises the complexity of the control flow. The inability to evaluate paths conclusively indicates a need to refine input constraints or further investigate symbolic execution limitations. Ultimately, this dataset offers deep insights into path feasibility without final verdict certainty.

The test generation summary reveals a comprehensive attempt by PathCrawler to traverse 19 total paths within the \texttt{Tritype} function. Of these, 11 were covered and 8 marked as infeasible. The success of achieving 100\% branch coverage suggests an effective instrumentation and exploration of all conditional branches in the function. However, none of the generated test cases resulted in a definitive "success" or "failure" outcome, indicating that path postconditions or functional assertions may not have been defined or interpreted robustly enough for classification. The discrepancy between path coverage and test verdicts signals potential limitations in either the underlying test oracle or symbolic constraint solving. Nevertheless, all test cases completed without runtime errors or crashes, which strengthens the confidence in static path feasibility. This situation is common in symbolic execution where certain paths are executable in theory but do not yield assertable functional correctness outcomes.

Several test cases (such as 3, 4, and 7) represent crucial boundary classifications in triangle identification logic. Test Case 3, returning 0, indicates that the input values do not form a valid triangle—corroborated by the path predicate involving non-equal sides and valid sum conditions. Test Cases 4 to 7 exhibit values returning 1 or 2, which correspond to isosceles and equilateral triangle scenarios. These variations affirm the tool’s capability to explore distinct triangle classifications based on the input side lengths. However, all cases have an "unknown" verdict. This suggests that while execution paths were valid, postconditions or specification assertions did not yield sufficient evidence for a conclusive outcome. It is also possible that comparison thresholds or floating-point equality checks introduced uncertainty. The tool's symbolic reasoning did cover a wide input range, showcasing strength in data diversity.

This selection emphasises the diversity in input ranges—both positive and negative, and also includes floating-point values. Negative and invalid inputs, such as in Case 1, correctly trace to a return of 3, denoting an error or invalid triangle. Equilateral triangles like in Case 7 return 2, consistent with the expected logic. The presence of precise floating values (e.g., 886.99) highlights the need for robust type handling in symbolic execution and comparison logic. The return values indicate that, functionally, the outputs align with expectations, even if PathCrawler’s analysis labels the verdict as unknown.

Examining the path predicates reveals detailed symbolic constraints applied during test generation. For example, Case 3 checks all triangle inequality constraints and non-equality of sides. Case 7's predicate confirms all sides are equal, ideal for verifying the equilateral branch of logic. These conditions provide assurance that symbolic execution adheres to logical expectations derived from triangle properties. However, predicates like those in Cases 1, 10, and 11 reflect invalid or boundary-breaking inputs—some involving negative sides. While such inputs yield predictable output values, their classification remains "unknown" due to a lack of explicit functional assertions. The predicates affirm that PathCrawler applies comprehensive constraint exploration but might benefit from clearer postcondition verification to move cases from “unknown” to “success” or “failure.”

The analysis indicates high internal code coverage but limited external verdict classification due to insufficient assertions or postcondition specifications. While symbolic execution succeeded in evaluating feasible and infeasible paths and fully traversing conditional logic, the lack of verdicts suggests an improvement opportunity in integrating output expectations with test oracles. PathCrawler’s utility in this context lies more in code path analysis and constraint checking than in final correctness validation. Future work may involve explicitly encoding expected behavior (e.g., via assertion annotations or return value comparisons) to convert "unknown" results into definitive outcomes. Despite these limitations, the tool has proven its capability in surfacing hidden paths and exercising diverse execution flows. Particularly in legacy or critical systems like \texttt{Tritype}, such analysis can prevent unanticipated bugs or logic flaws through exhaustive test path enumeration.

\section*{Analysis of Verification on \texttt{baseline\_Example1-Tritype.c}}

The baseline prompt is available on page 20-21, \cite{Granberry2025a}.
The verification of \texttt{baseline\_Example1-Tritype.c} using Frama-C and four SMT solvers (Alt-Ergo, Z3, CVC4, and CVC5) revealed a consistent pattern. With minimal ACSL specification, only two implicit verification goals—termination and unreachability—were generated and successfully verified by all four provers. This indicates that the function is syntactically well-formed and does not exhibit trivial issues like infinite loops or unreachable code paths. However, due to the absence of user-defined postconditions or preconditions, the analysis provides limited insight into functional correctness. The warning about the missing \texttt{assigns} clause suggests that memory side-effects are not specified, potentially causing inaccurate assumptions for callers. Similarly, the absence of RTE (Run-Time Error) guards indicates that common runtime errors like overflows or division by zero are not being verified. While the results demonstrate soundness at a structural level, the lack of deep specification significantly limits the utility of the verification. All provers perform equally under these trivial conditions.

\section*{Comparison of Prover Results on \texttt{baseline\_Example1-Tritype.c}}

\begin{table}[h]
\centering
\caption{Prover Results for \texttt{baseline\_Example1-Tritype.c}}
\begin{tabular}{@{}lccc@{}}
\toprule
\textbf{Prover} & \textbf{Total Goals} & \textbf{Proved} & \textbf{Notes} \\ \midrule
Alt-Ergo & 2 & 2 & All default goals proved \\
Z3       & 2 & 2 & All default goals proved \\
CVC4     & 2 & 2 & All default goals proved \\
CVC5     & 2 & 2 & All default goals proved \\
\bottomrule
\end{tabular}
\end{table}

\subsection*{Analysis of Verification on pathcrawler\_augmented\_Example1-Tritype.c}

The pathcrawler augmented prompt is available on page 21-22, \cite{Granberry2025a} while EVA augmented prompt is available on page 24, \cite{Granberry2025a}.
The \texttt{pathcrawler\_augmented\_Example1-Tritype.c} file, enriched with detailed ACSL annotations, exhibits significantly different verification behavior. A total of 20 goals were generated, including 18 based on user-specified preconditions and postconditions, plus the standard termination and unreachability checks. The analysis reveals a distinct divide in prover effectiveness. While all provers verified termination and basic logic, their ability to handle complex ensures clauses varied. Z3 and CVC5 each failed to prove seven goals—Z3 due to timeouts and CVC4 due to unknown statuses—highlighting their struggles with intricate logical paths and case distinctions. Alt-Ergo and CVC5 fared slightly better, with only five unverified goals each. Notably, the most complex properties, such as correct classification of triangle types (Scalene, Isosceles, Equilateral) and handling of inequality rules, were consistently problematic across all provers. This reflects the challenges solvers face when dealing with disjunction-heavy logic or subtle arithmetic constraints embedded in functional specifications.

\subsection*{Comparison of Prover Results on pathcrawler\_augmented\_Example1-Tritype.c}

\begin{table}[h]
\centering
\caption{Prover Results for \texttt{pathcrawler\_augmented\_Example1-Tritype.c}}
\begin{tabular}{@{}lcccc@{}}
\toprule
\textbf{Prover} & \textbf{Total Goals} & \textbf{Proved} & \textbf{Failed Type} & \textbf{Failed Count} \\ \midrule
Z3       & 20 & 13 & Timeout  & 7 \\
Alt-Ergo & 20 & 15 & Timeout  & 5 \\
CVC4     & 20 & 13 & Unknown  & 7 \\
CVC5     & 20 & 15 & Timeout  & 5 \\
\bottomrule
\end{tabular}
\end{table}

\section{Performance of provers on 36 Examples of Frama-C Tutorial}
\label{sec:exectime_analysis}
\subsection*{RTE Tool Output Summary}

\scriptsize
\begin{longtable}{p{3.5cm} >{\centering\arraybackslash}p{0.8cm} >{\centering\arraybackslash}p{1.3cm} >{\centering\arraybackslash}p{0.7cm} >{\centering\arraybackslash}p{0.9cm} >{\centering\arraybackslash}p{0.9cm} >{\centering\arraybackslash}p{0.9cm} >{\centering\arraybackslash}p{1cm} >{\centering\arraybackslash}p{1.8cm}}
\toprule
\textbf{C File} & \textbf{Goals} & \textbf{Proved} & \textbf{Qed} & \textbf{Timeout} & \textbf{Term.} & \textbf{Unreach.} & \textbf{Alt-Ergo} & \textbf{Assigns Missing} \\
\midrule
\endfirsthead

\toprule
\textbf{C File} & \textbf{Goals} & \textbf{Proved} & \textbf{Qed} & \textbf{Timeout} & \textbf{Term.} & \textbf{Unreach.} & \textbf{Alt-Ergo} & \textbf{Assigns Missing} \\
\midrule
\endhead

01-abs-0.c & 1 & 2 / 3 & 0 & 1 & 1 & 1 & 0 & Yes \\
01-abs-1.c & 2 & 3 / 4 & 1 & 1 & 1 & 1 & 0 & Yes \\
01-abs-2.c & 4 & 6 / 6 & 4 & 0 & 1 & 1 & 0 & No \\
01-abs-3.c & 5 & 7 / 7 & 5 & 0 & 1 & 1 & 0 & No \\
02-max-0.c & 2 & 4 / 4 & 2 & 0 & 1 & 1 & 0 & Yes \\
02-max-1.c & 5 & 5 / 7 & 3 & 2 & 1 & 1 & 0 & Yes \\
02-max-2.c & 1 & 3 / 3 & 0 & 0 & 1 & 1 & 1 & Yes \\
02-max-3.c & 6 & 7 / 8 & 4 & 1 & 1 & 1 & 1 & Yes \\
02-max-4.c & 8 & 10 / 10 & 7 & 0 & 1 & 1 & 1 & Yes \\
03-max\_ptr-0.c & 4 & 4 / 6 & 2 & 2 & 1 & 1 & 0 & Yes \\
03-max\_ptr-1.c & 6 & 6 / 8 & 4 & 2 & 1 & 1 & 0 & Yes \\
03-max\_ptr-2.c & 6 & 8 / 8 & 4 & 0 & 1 & 1 & 2 & Yes \\
03-max\_ptr-3.c & 4 & 6 / 6 & 3 & 0 & 1 & 1 & 1 & Yes \\
03-max\_ptr-4.c & 8 & 10 / 10 & 6 & 0 & 1 & 1 & 2 & No \\
04-incr\_a\_by\_b-0.c & 5 & 3 / 7 & 0 & 4 & 1 & 1 & 1 & Yes \\
04-incr\_a\_by\_b-1.c & 7 & 9 / 9 & 4 & 0 & 1 & 1 & 3 & No \\
04-incr\_a\_by\_b-fail.c & 7 & 8 / 9 & 4 & 1 & 1 & 1 & 2 & No \\
04-swap-0.c & 6 & 4 / 8 & 2 & 4 & 1 & 1 & 0 & Yes \\
04-swap-1.c & 12 & 14 / 14 & 9 & 0 & 1 & 1 & 3 & Yes \\
05-abs-0.c & 1 & 2 / 3 & 0 & 1 & 1 & 1 & 0 & Yes \\
05-abs-1.c & 7 & 9 / 9 & 7 & 0 & 1 & 1 & 0 & No \\
05-abs-2.c & 6 & 8 / 8 & 6 & 0 & 1 & 1 & 0 & No \\
06-max\_abs-0.c & 2 & 2 / 2 & 2 & 0 & -- & -- & 0 & Yes \\
06-max\_abs-1.c & 13 & 11 / 13 & 9 & 2 & -- & -- & 2 & Yes \\
06-max\_abs-2.c & 13 & 12 / 13 & 11 & 1 & -- & -- & 1 & Yes \\
06-max\_abs-3.c & 13 & 13 / 13 & 11 & 0 & -- & -- & 2 & Yes \\
07-reset\_array-0.c & 3 & 2 / 4 & 1 & 2 & -- & 1 & 0 & Yes \\
07-reset\_array-1.c & 13 & 15 / 15 & 9 & 0 & 1 & 1 & 4 & No \\
08-binary\_search-1.c & 27 & 29 / 29 & 13 & 0 & 1 & 1 & 14 & No \\
09-sqrt-0.c & 6 & 4 / 7 & 2 & 3 & -- & 1 & 1 & Yes \\
\bottomrule
\end{longtable}

\subsection*{Z3 Prover Output Summary}

\scriptsize
\begin{longtable}{p{3.5cm} >{\centering\arraybackslash}p{0.8cm} >{\centering\arraybackslash}p{1.3cm} >{\centering\arraybackslash}p{0.7cm} >{\centering\arraybackslash}p{0.9cm} >{\centering\arraybackslash}p{0.9cm} >{\centering\arraybackslash}p{0.9cm} >{\centering\arraybackslash}p{1cm} >{\centering\arraybackslash}p{1.8cm}}
\toprule
\textbf{C File} & \textbf{Goals} & \textbf{Proved} & \textbf{Qed} & \textbf{Timeout} & \textbf{Term.} & \textbf{Unreach.} & \textbf{Z3} & \textbf{Assigns Missing} \\
\midrule
\endfirsthead

\toprule
\textbf{C File} & \textbf{Goals} & \textbf{Proved} & \textbf{Qed} & \textbf{Timeout} & \textbf{Term.} & \textbf{Unreach.} & \textbf{Z3} & \textbf{Assigns Missing} \\
\midrule
\endhead

01-abs-0.c & 0 & 2 / 2 & 1 & 1 & 0 & 0 & 0 & Yes \\
01-abs-1.c & 1 & 3 / 3 & 1 & 1 & 1 & 0 & 0 & Yes \\
01-abs-2.c & 3 & 5 / 5 & 1 & 1 & 3 & 0 & 0 & No \\
01-abs-3.c & 4 & 6 / 6 & 1 & 1 & 4 & 0 & 0 & No \\
02-max-0.c & 2 & 4 / 4 & 1 & 1 & 2 & 0 & 0 & Yes \\
02-max-1.c & 5 & 5 / 7 & 1 & 1 & 3 & 0 & 2 & Yes \\
02-max-2.c & 1 & 3 / 3 & 1 & 1 & 0 & 1 & 0 & Yes \\
02-max-3.c & 6 & 7 / 8 & 1 & 1 & 4 & 1 & 1 & Yes \\
02-max-4.c & 8 & 10 / 10 & 1 & 1 & 7 & 1 & 0 & Yes \\
03-max\_ptr-0.c & 0 & 2 / 2 & 1 & 1 & 0 & 0 & 0 & Yes \\
03-max\_ptr-1.c & 2 & 4 / 4 & 1 & 1 & 2 & 0 & 0 & Yes \\
03-max\_ptr-2.c & 2 & 4 / 4 & 1 & 1 & 2 & 0 & 0 & Yes \\
03-max\_ptr-3.c & 2 & 4 / 4 & 1 & 1 & 1 & 1 & 0 & Yes \\
03-max\_ptr-4.c & 4 & 6 / 6 & 1 & 1 & 4 & 0 & 0 & No \\
04-incr\_a\_by\_b-0.c & 0 & 2 / 2 & 1 & 1 & 0 & 0 & 0 & Yes \\
04-incr\_a\_by\_b-1.c & 2 & 4 / 4 & 1 & 1 & 1 & 1 & 0 & No \\
04-incr\_a\_by\_b-fail.c & 2 & 3 / 4 & 1 & 1 & 1 & 0 & 1 & No \\
04-swap-0.c & 2 & 4 / 4 & 1 & 1 & 2 & 0 & 0 & Yes \\
04-swap-1.c & 8 & 10 / 10 & 1 & 1 & 7 & 1 & 0 & Yes \\
05-abs-0.c & 0 & 2 / 2 & 1 & 1 & 0 & 0 & 0 & Yes \\
05-abs-1.c & 6 & 8 / 8 & 1 & 1 & 6 & 0 & 0 & No \\
05-abs-2.c & 5 & 7 / 7 & 1 & 1 & 5 & 0 & 0 & No \\
06-max\_abs-0.c & 2 & 2 / 2 & 0 & 0 & 2 & 0 & 0 & Yes \\
06-max\_abs-1.c & 13 & 11 / 13 & 0 & 0 & 9 & 2 & 2 & No \\
06-max\_abs-2.c & 13 & 12 / 13 & 0 & 0 & 11 & 1 & 1 & No \\
06-max\_abs-3.c & 13 & 13 / 13 & 0 & 0 & 11 & 2 & 0 & No \\
07-reset\_array-0.c & 1 & 1 / 2 & 0 & 1 & 0 & 0 & 1 & Yes \\
07-reset\_array-1.c & 11 & 13 / 13 & 1 & 1 & 8 & 3 & 0 & No \\
08-binary\_search-1.c & 18 & 18 / 20 & 1 & 1 & 11 & 5 & 2 & No \\
\bottomrule
\end{longtable}

\subsection*{CVC4 Prover Output Summary}

\scriptsize
\begin{longtable}{p{3.5cm} >{\centering\arraybackslash}p{0.8cm} >{\centering\arraybackslash}p{1.3cm} >{\centering\arraybackslash}p{0.7cm} >{\centering\arraybackslash}p{0.9cm} >{\centering\arraybackslash}p{0.9cm} >{\centering\arraybackslash}p{0.9cm} >{\centering\arraybackslash}p{1cm} >{\centering\arraybackslash}p{1.8cm}}
\toprule
\textbf{C File} & \textbf{Goals} & \textbf{Proved} & \textbf{Qed} & \textbf{Timeout} & \textbf{Term.} & \textbf{Unreach.} & \textbf{CVC4} & \textbf{Assigns Missing} \\
\midrule
\endfirsthead

\toprule
\textbf{C File} & \textbf{Goals} & \textbf{Proved} & \textbf{Qed} & \textbf{Timeout} & \textbf{Term.} & \textbf{Unreach.} & \textbf{CVC4} & \textbf{Assigns Missing} \\
\midrule
\endhead

01-abs-0.c & 0 & 2 / 2 & 1 & 1 & 0 & 0 & 0 & Yes \\
01-abs-1.c & 1 & 3 / 3 & 1 & 1 & 1 & 0 & 0 & Yes \\
01-abs-2.c & 3 & 5 / 5 & 1 & 1 & 3 & 0 & 0 & No \\
01-abs-3.c & 4 & 6 / 6 & 1 & 1 & 4 & 0 & 0 & No \\
02-max-0.c & 2 & 4 / 4 & 1 & 1 & 2 & 0 & 0 & Yes \\
02-max-1.c & 5 & 5 / 7 & 1 & 1 & 3 & 0 & 2 & Yes \\
02-max-2.c & 1 & 3 / 3 & 1 & 1 & 0 & 1 & 0 & Yes \\
02-max-3.c & 6 & 7 / 8 & 1 & 1 & 4 & 1 & 1 & Yes \\
02-max-4.c & 8 & 10 / 10 & 1 & 1 & 7 & 1 & 0 & Yes \\
03-max\_ptr-0.c & 0 & 2 / 2 & 1 & 1 & 0 & 0 & 0 & Yes \\
03-max\_ptr-1.c & 2 & 4 / 4 & 1 & 1 & 2 & 0 & 0 & Yes \\
03-max\_ptr-2.c & 2 & 4 / 4 & 1 & 1 & 2 & 0 & 0 & Yes \\
03-max\_ptr-3.c & 2 & 4 / 4 & 1 & 1 & 1 & 1 & 0 & Yes \\
03-max\_ptr-4.c & 4 & 6 / 6 & 1 & 1 & 4 & 0 & 0 & No \\
04-incr\_a\_by\_b-0.c & 0 & 2 / 2 & 1 & 1 & 0 & 0 & 0 & Yes \\
04-incr\_a\_by\_b-1.c & 2 & 4 / 4 & 1 & 1 & 1 & 1 & 0 & No \\
04-incr\_a\_by\_b-fail.c & 2 & 3 / 4 & 1 & 1 & 1 & 0 & 1 & No \\
04-swap-0.c & 2 & 4 / 4 & 1 & 1 & 2 & 0 & 0 & Yes \\
04-swap-1.c & 8 & 10 / 10 & 1 & 1 & 7 & 1 & 0 & Yes \\
05-abs-0.c & 0 & 2 / 2 & 1 & 1 & 0 & 0 & 0 & Yes \\
05-abs-1.c & 6 & 8 / 8 & 1 & 1 & 6 & 0 & 0 & No \\
05-abs-2.c & 5 & 7 / 7 & 1 & 1 & 5 & 0 & 0 & No \\
06-max\_abs-0.c & 2 & 2 / 2 & 0 & 0 & 2 & 0 & 0 & Yes \\
06-max\_abs-1.c & 13 & 11 / 13 & 0 & 0 & 9 & 2 & 2 & No \\
06-max\_abs-2.c & 13 & 12 / 13 & 0 & 0 & 11 & 1 & 1 & No \\
06-max\_abs-3.c & 13 & 13 / 13 & 0 & 0 & 11 & 2 & 0 & No \\
07-reset\_array-0.c & 1 & 1 / 2 & 0 & 1 & 0 & 0 & 1 & Yes \\
07-reset\_array-1.c & 11 & 13 / 13 & 1 & 1 & 8 & 3 & 0 & No \\
08-binary\_search-1.c & 18 & 16 / 20 & 1 & 1 & 11 & 3 & 4 & No \\
\bottomrule
\end{longtable}

\subsection*{CVC5 Prover Output Summary}

\scriptsize
\begin{longtable}{p{3.5cm} >{\centering\arraybackslash}p{0.8cm} >{\centering\arraybackslash}p{1.3cm} >{\centering\arraybackslash}p{0.7cm} >{\centering\arraybackslash}p{0.9cm} >{\centering\arraybackslash}p{0.9cm} >{\centering\arraybackslash}p{0.9cm} >{\centering\arraybackslash}p{1cm} >{\centering\arraybackslash}p{1.8cm}}
\toprule
\textbf{C File} & \textbf{Goals} & \textbf{Proved} & \textbf{Qed} & \textbf{Timeout} & \textbf{Term.} & \textbf{Unreach.} & \textbf{CVC5} & \textbf{Assigns Missing} \\
\midrule
\endfirsthead

\toprule
\textbf{C File} & \textbf{Goals} & \textbf{Proved} & \textbf{Qed} & \textbf{Timeout} & \textbf{Term.} & \textbf{Unreach.} & \textbf{CVC5} & \textbf{Assigns Missing} \\
\midrule
\endhead

01-abs-0.c & 0 & 2 / 2 & 1 & 1 & 0 & 0 & 0 & Yes \\
01-abs-1.c & 1 & 3 / 3 & 1 & 1 & 1 & 0 & 0 & Yes \\
01-abs-2.c & 3 & 5 / 5 & 1 & 1 & 3 & 0 & 0 & No \\
01-abs-3.c & 4 & 6 / 6 & 1 & 1 & 4 & 0 & 0 & No \\
02-max-0.c & 2 & 4 / 4 & 1 & 1 & 2 & 0 & 0 & Yes \\
02-max-1.c & 5 & 5 / 7 & 1 & 1 & 3 & 0 & 2 & Yes \\
02-max-2.c & 1 & 3 / 3 & 1 & 1 & 0 & 1 & 0 & Yes \\
02-max-3.c & 6 & 7 / 8 & 1 & 1 & 4 & 1 & 1 & Yes \\
02-max-4.c & 8 & 10 / 10 & 1 & 1 & 7 & 1 & 0 & Yes \\
03-max\_ptr-0.c & 0 & 2 / 2 & 1 & 1 & 0 & 0 & 0 & Yes \\
03-max\_ptr-1.c & 2 & 4 / 4 & 1 & 1 & 2 & 0 & 0 & Yes \\
03-max\_ptr-2.c & 2 & 4 / 4 & 1 & 1 & 2 & 0 & 0 & Yes \\
03-max\_ptr-3.c & 2 & 4 / 4 & 1 & 1 & 1 & 1 & 0 & Yes \\
03-max\_ptr-4.c & 4 & 6 / 6 & 1 & 1 & 4 & 0 & 0 & No \\
04-incr\_a\_by\_b-0.c & 0 & 2 / 2 & 1 & 1 & 0 & 0 & 0 & Yes \\
04-incr\_a\_by\_b-1.c & 2 & 4 / 4 & 1 & 1 & 1 & 1 & 0 & No \\
04-incr\_a\_by\_b-fail.c & 2 & 3 / 4 & 1 & 1 & 1 & 0 & 1 & No \\
04-swap-0.c & 2 & 4 / 4 & 1 & 1 & 2 & 0 & 0 & Yes \\
04-swap-1.c & 8 & 10 / 10 & 1 & 1 & 7 & 1 & 0 & Yes \\
05-abs-0.c & 0 & 2 / 2 & 1 & 1 & 0 & 0 & 0 & Yes \\
05-abs-1.c & 6 & 8 / 8 & 1 & 1 & 6 & 0 & 0 & No \\
05-abs-2.c & 5 & 7 / 7 & 1 & 1 & 5 & 0 & 0 & No \\
06-max\_abs-0.c & 2 & 2 / 2 & 0 & 0 & 2 & 0 & 0 & Yes \\
06-max\_abs-1.c & 13 & 11 / 13 & 0 & 0 & 9 & 2 & 2 & No \\
06-max\_abs-2.c & 13 & 12 / 13 & 0 & 0 & 11 & 1 & 1 & No \\
06-max\_abs-3.c & 13 & 13 / 13 & 0 & 0 & 11 & 2 & 0 & No \\
07-reset\_array-0.c & 1 & 1 / 2 & 0 & 1 & 0 & 0 & 1 & Yes \\
07-reset\_array-1.c & 11 & 13 / 13 & 1 & 1 & 8 & 3 & 0 & No \\
08-binary\_search-1.c & 18 & 16 / 20 & 1 & 1 & 11 & 3 & 4 & No \\
\bottomrule
\end{longtable}

\subsection*{Execution Times}
\label{app:exectime_analysis}
\scriptsize
\begin{longtable}{p{3.5cm} >{\centering\arraybackslash}p{2.5cm} >{\centering\arraybackslash}p{2cm} >{\centering\arraybackslash}p{2.5cm} >{\centering\arraybackslash}p{2.5cm}}
\toprule
\textbf{File Name} & \textbf{Alt-ergo Time(s)} & \textbf{Z3 Time(s)} & \textbf{CVC4 Time(s)} & \textbf{CVC5 Time(s)} \\
\midrule
\endfirsthead

\toprule
\textbf{File Name} & \textbf{Alt-ergo Time(s)} & \textbf{Z3 Time(s)} & \textbf{CVC4 Time(s)} & \textbf{CVC5 Time(s)} \\
\midrule
\endhead

01-abs-0.c & 0.01 & 0.02 & 0.02 & 0.01 \\
01-abs-1.c & 0.02 & 0.04 & 0.02 & 0.001 \\
01-abs-2.c & 0.05 & 0.06 & 0.05 & 0.003 \\
01-abs-3.c & 0.04 & 0.03 & 0.03 & 0.004 \\
02-max-0.c & 0.02 & 0.02 & 0.01 & 0.002 \\
02-max-1.c & 0.07 & 0.06 & 0.05 & 0.004 \\
02-max-2.c & 0.08 & 0.07 & 0.05 & 0.002 \\
02-max-3.c & 0.12 & 0.14 & 0.13 & 0.004 \\
02-max-4.c & 0.05 & 0.06 & 0.05 & 0.007 \\
03-max\_ptr-0.c & 0.02 & 0.03 & 0.02 & 0.001 \\
03-max\_ptr-1.c & 0.03 & 0.04 & 0.03 & 0.002 \\
03-max\_ptr-2.c & 0.06 & 0.06 & 0.06 & 0.002 \\
03-max\_ptr-3.c & 0.09 & 0.09 & 0.07 & 0.003 \\
03-max\_ptr-4.c & 0.07 & 0.07 & 0.08 & 0.004 \\
04-incr\_a\_by\_b-0.c & 0.03 & 0.04 & 0.04 & 0.002 \\
04-incr\_a\_by\_b-1.c & 0.02 & 0.03 & 0.01 & 0.003 \\
04-incr\_a\_by\_b-fail.c & 0.03 & 0.04 & 0.04 & 0.002 \\
04-swap-0.c & 0.01 & 0.02 & 0.02 & 0.002 \\
04-swap-1.c & 0.04 & 0.05 & 0.03 & 0.004 \\
05-abs-0.c & 0.05 & 0.07 & 0.06 & 0.004 \\
05-abs-1.c & 0.07 & 0.08 & 0.07 & 0.006 \\
05-abs-2.c & 0.09 & 0.10 & 0.09 & 0.005 \\
06-max\_abs-0.c & 0.03 & 0.04 & 0.03 & 0.002 \\
06-max\_abs-1.c & 0.11 & 0.12 & 0.11 & 0.009 \\
06-max\_abs-2.c & 0.14 & 0.15 & 0.14 & 0.011 \\
06-max\_abs-3.c & 0.13 & 0.14 & 0.13 & 0.011 \\
07-reset\_array-0.c & 0.02 & 0.03 & 0.02 & 0.002 \\
07-reset\_array-1.c & 0.08 & 0.10 & 0.09 & 0.008 \\
08-binary\_search-1.c & 0.25 & 0.28 & 0.30 & 0.064 \\
\bottomrule
\caption{Execution times of Alt-Ergo, Z3, CVC4 and CVC5 on C files}
\label{table:execution_times}
\end{longtable}

\begin{figure}[ht]
    \centering
		\includegraphics[width=\linewidth]{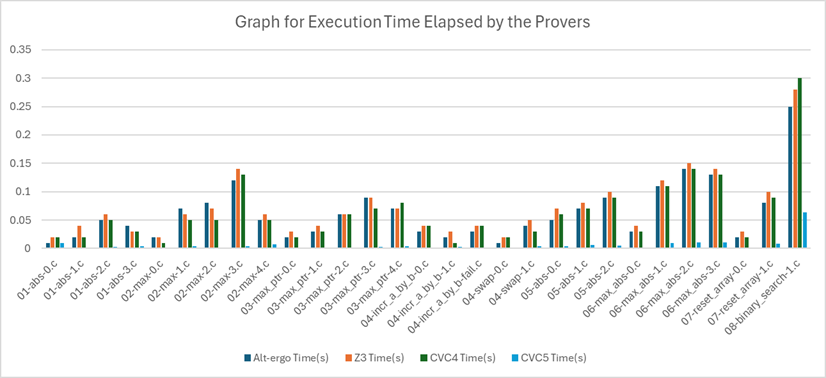}
    \caption{Graph of Execution Times of Alt-Ergo, Z3, CVC4 and CVC5}
    \label{fig:execution_times}
\end{figure}

{\normalsize
\section{Description of Challenges and Future Directions}
\label{app:challenges_future}

\noindent\textbf{C1: Semantic Ambiguity in Natural Language} - The ambiguity introduced in software requirements through the use of natural language is evident. The context-dependent terms, implicit references, and domain-specific jargon complicate accurate interpretation of requirements for automatic specification generation. The use of LLMs helps address this challenge but mis-interpretation of user intent still leads to incorrect outputs. Removing ambiguity requires structured domain knowledge aligned with external ontologies. As the literature synthesis suggests, improved models and effective human-in-the- loop strategies increase the accuracy in generating correct and verifiable specifications. 

\noindent\textbf{C2: Lack of Ground Truth Datasets} -  public datasets involving industry based software is difficult  to locate, arguably due to the sensitivity of software pieces and underlying IP rights. This scarcity of standard datasets comprising of natural language requirements together with formal logic poses an obstacle in the much required steps of model training, evaluation, and reproducibility (core components of achieving better accuracy in mainstream NLP tasks). It also involves a scalability challenge. Advancing the field requires high-quality, and annotated datasets that span varied industries and requirement types. The synthesised  datasets partially solve the problem. The availability of high- quality, curated and annotated standardised formal specification datasets involving varied industries and requirement types would  push the boundary in this domain. These datasets should capture real-world software complexity and domain diversity.

\noindent\textbf{C3: Tool Interoperability} - There is a significant interest in developing tool-chains to integrate formal verification tools for verifying critical software from the last two decades. Formal verification process rely on diverse tools with incompatible formats and limited integration capabilities, which  makes end-to-end automation difficult. The process of integrating formal verification tools involves manual or semi-automatic intervention. The process becomes difficult in the absence of shared standards or modular pipelines. Achieving seamless and low-effort tool interoperability requires standardised interfaces and ontologies.

\noindent\textbf{C4: Traceability Across Artefact Lifecycles} - A typical software development life-cycle involves multiple phases. Maintaining traceability and alignment throughout the development cycle is the core requirement. Achieving traceability manually can compromise system integrity. Effective traceability links rationale, changes and dependencies across software artefacts i.e. text, models, code, and tests. LLMs can assist in impact analysis to maintain traceability, but imposes a challenge of collaborative, explainable, and context-aware evolution process.

\noindent\textbf{C5: Explainability and User Trust} - For LLM-generated formal specifications to be trusted, users must understand how outputs were derived and whether they reflect intended meaning. Current models offer limited transparency, often lacking rationale or input attribution. Addressing this involves adding annotations in the implementation and can potentially be resolved through use of LLMs. But LLM generated annotations can be under-developed or involve quality check using a human-in-the-loop, especially ensuring trust in safety-critical systems and impose the challenge of deploying inspection, refinement, and human oversight.

\begin{figure}[ht]
\centering
\scalebox{0.8}{
\begin{tikzpicture}[
  challenge/.style={draw=blue!60!black, fill=blue!10, thick, text width=6cm, align=left, minimum height=1.2cm},
  future/.style={draw=green!60!black, fill=green!10, thick, text width=6cm, align=left, minimum height=1.2cm},
  arrow/.style={-{hooks}, thick},
  node distance=0.4cm and 1cm
]
\node[challenge] (c1) {C1: Semantic Ambiguity};
\node[challenge, below=of c1] (c2) {C2: Lack of Ground Truth Datasets};
\node[challenge, below=of c2] (c3) {C3: Tool Interoperability};
\node[challenge, below=of c3] (c4) {C4: Traceability Across Artefact Lifecycles};
\node[challenge, below=of c4] (c5) {C5: Explainability and User Trust};

\node[future, right=of c1] (f1) {F1: Human-in-the-loop Formalisation};
\node[future, below=of f1] (f2) {F2: Multi-modal Artefact Alignment};
\node[future, below=of f2] (f3) {F3: Standardised Benchmarks};
\node[future, below=of f3] (f4) {F4: Neuro-symbolic Reasoning};
\node[future, below=of f4] (f5) {F5: Interactive Traceability Tools};

\draw[->, thick, >=stealth] (f1.west) -- (c1.east);
\draw[->, thick, >=stealth] (f1.west) -- (c5.east);
\draw[->, thick, >=stealth] (f2.west) -- (c1.east);
\draw[->, thick, >=stealth] (f2.west) -- (c4.east);
\draw[->, thick, >=stealth] (f3.west) -- (c2.east);
\draw[->, thick, >=stealth] (f3.west) -- (c5.east);
\draw[->, thick, >=stealth] (f4.west) -- (c1.east);
\draw[->, thick, >=stealth] (f4.west) -- (c2.east);
\draw[->, thick, >=stealth] (f4.west) -- (c3.east);
\draw[->, thick, >=stealth] (f4.west) -- (c5.east);
\draw[->, thick, >=stealth] (f5.west) -- (c4.east);
\draw[->, thick, >=stealth] (f5.west) -- (c5.east);

\end{tikzpicture}
}
\caption{Mapping between challenges (C1 – C5) and future research  (F1- F5) directions in LLM-based formalisation}
\label{fig:challenges_future}
\end{figure}

\noindent\textbf{F1: Human-in-the-loop Formalisation} - In this approach, the formalisation is semi-automated and guided by a domain expert. LLMs can assist in proposing logic, trace links, or suggest refinements which can be approved or revised by the users. The process ensures less ambiguity, improved accuracy and increased user trust. It enhances learning as well because improvement in model outputs is ensured through interaction. This process can be facilitated by better visualisation, iteration, and feedback support. This approach facilitates adopting new methodologies as human judgment remains central to critical decisions.

\noindent\textbf{F2: Multi-modal Artefact Alignment} - Software requirements can be documented through a combination of raised through  text, diagrams, tables, and spreedsheets, depending on the methodology adopted by the developers. Alignment between these can reduce ambiguity and lead to better formal specifications. Multi-modal support in the tools  should be available and can be achieved through representation learning and semantic matching. LLMs having structural and visual modalities can improve context interpretation, resulting in more comprehensive models. Formalisation pipelines and redundant artefact types can model real-world complexity, impacting the correctness and reliability of generated specifications.

\noindent\textbf{F3: Standardised Benchmarks} - The future direction of developing standardised benchmark has one-to-one mapping to identified challenge C2. Having standardised benchmarks to simulate and compare performance is a well-established research area e.g. \cite{systems13070567, Gregory2020, PROMISE2007DataSet, RETRONET2018}. Advancing the field requires high-quality, and annotated datasets that span varied industries and requirement types. Though the synthesised datasets have partially solved the problem, the availability of high-quality and annotated standardised formal specifications datasets remains limited. However, such datasets involving varied industries and requirement types can push the boundary in this domain. These datasets should capture real-world software complexity and domain diversity. As the field matures, such resources will become an essential component for progress.

\noindent\textbf{F4: Neuro-symbolic Reasoning} - Neuro-symbolic systems combine the adaptability of LLMs with the precision of symbolic reasoning. In these setups, neural networks suggest possible specifications While logic-based tools check or refine them. This helps improve consistency, enforce constraints, and verify correctness. Symbolic features like type rules or domain-specific logic can also guide the learning process. While building such systems is complex, the hybrid method offers the potential of lowering hallucinations and boosting clarity. Ongoing research could lead to reliable, interpretable models that merge statistical learning with formal methods.

\noindent\textbf{F5: Interactive Traceability Tools} - Improving traceability of software requirements through interactive user-friendly tools is a well-established area of requirements engineering. Traceability needs to be built into the formalisation process from the start and interactive tools can link requirements to models, tests, and verification results. This supports debugging, system updates, and audits. Features like visual navigation, filtering, and tagging help teams track changes and spot dependencies. LLMs can help by proposing trace links, flagging inconsistencies, or explaining revisions but users must stay in control of key decisions. These tools should work within development environments and support team collaboration. When traceability is clear and usable, it boosts system transparency and helps ensure compliance—especially in regulated fields like aerospace and healthcare. Traceability tools like EARS, FRET and DOORS are industry-proven. Industry-based case studies are a valuable avenue for future work in this field. 

\end{document}